%% file: main.tex
\documentclass[conference]{IEEEtran}
\IEEEoverridecommandlockouts
\usepackage{cite}
\usepackage{url} 
\usepackage{amsmath,amssymb,amsfonts}
\usepackage[noend]{algorithmic}
\usepackage{graphicx}
\usepackage{subcaption} 
\usepackage{textcomp}
\usepackage{xcolor}
\usepackage{quantikz}
\usepackage{tikz}
\usepackage{amsmath}
\usepackage{mathrsfs, comment}
\usepackage{array}

\usepackage{adjustbox}
\usepackage{color,soul}

\newcommand{\sol}{FitCut}

\newcommand{\objOne}{$Obj_{nc}$}
\newcommand{\objTwo}{$Obj_{ru}$}

\usepackage[tt=false, type1=true]{libertine}
\usepackage[utf8]{inputenc}

\usepackage{algorithm,multirow}
\begin{document}

\title{\huge Scalable Circuit Cutting and Scheduling in a Resource-constrained and Distributed Quantum System
}

\author{

Shuwen Kan\textsuperscript{\rm 1},
Zefan Du\textsuperscript{\rm 1},
Miguel Palma\textsuperscript{\rm 1},\\
Samuel A Stein\textsuperscript{\rm 2},
Chenxu Liu\textsuperscript{\rm 2},
Wenqi Wei\textsuperscript{\rm 1},
Juntao Chen\textsuperscript{\rm 1},
Ang Li\textsuperscript{\rm 2},
and Ying Mao\textsuperscript{\rm 1}
\\

 \textsuperscript{\rm 1} Computer and Information Science Department, Fordham University, \\ \{sk107, zdu19, mip2, wwei23, jchen504, ymao41\}@fordham.edu\\
    \textsuperscript{\rm 2}Pacific Northwest National Laboratory (PNNL), \{samuel.stein, chenxu.liu, ang.li\}@pnnl.gov\\
}

\maketitle

\begin{abstract}

Despite quantum computing's rapid development, current systems remain limited in practical applications due to their limited qubit count and quality. Various technologies, such as superconducting, trapped ions, and neutral atom quantum computing technologies are progressing towards a fault tolerant era, however they all face a diverse set of challenges in scalability and control. Recent efforts have focused on multi-node quantum systems that connect multiple smaller quantum devices to execute larger circuits. Future demonstrations hope to use quantum channels to couple systems, however current demonstrations can leverage classical communication with circuit cutting techniques. This involves cutting large circuits into smaller subcircuits and reconstructing them post-execution. However, existing cutting methods are hindered by lengthy search times as the number of qubits and gates increases. Additionally, they often fail to effectively utilize the resources of various worker configurations in a multi-node system. To address these challenges, we introduce \sol, a novel approach that transforms quantum circuits into weighted graphs and utilizes a community-based, bottom-up approach to cut circuits according to resource constraints, e.g., qubit counts, on each worker. \sol~also includes a scheduling algorithm that optimizes resource utilization across workers. Implemented with Qiskit and evaluated extensively, \sol~significantly outperforms the Qiskit Circuit Knitting Toolbox, reducing time costs by factors ranging from 3 to 2000 and improving resource utilization rates by up to 3.88 times on the worker side, achieving a system-wide improvement of 2.86 times.

\end{abstract}


\input{introduction}

\input{relatedworks}
\input{background}

\input{solution-design}

\input{algorithm}
\input{evaluation}

\input{conclusion}

\bibliographystyle{IEEEtran}
\bibliography{references}

\end{document}

%% file: introduction.tex
\section{Introduction}
The potential of quantum-based computing systems continues to garner significant interest from both industry and academia, with many eager to achieve quantum speedup for different applications, such as innovations in quantum machine learning~\cite{stein2021hybrid, stein2022qucnn, stein2022quclassi, valdez2023review, stein2021qugan, higham2023quantum, l2024quantum, d2023distributed}, quantum search~\cite{mu2022iterative, pokharel2024better, park2023quantum, seidel2023automatic}, quantum chemistry~\cite{simons2023quantum, sheng2023quantum, karabacak2023novel} and quantum drug discovery~\cite{ wang2023recent, santagati2024drug, posenitskiy2023trexio} and understanding of quantum systems through different techniques, such as noise characterization~\cite{dahlhauser2024benchmarking, baheri2022pinpointing, hamilton2020scalable} and visualization~\cite{ruan2023venus, santurkarinvestigation, ruan2022vacsen, barthe2023bloch,ruan2023quantumeyes}. 
Despite rapid advancements in quantum computing, practical applications remain elusive due to significant challenges~\cite{de2021materials}. 
Technologies like superconducting qubits~\cite{ibmQuantum, rigettiQuantumComputing}, trapped ions~\cite{ionqIonQTrapped, quantinuumAcceleratingQuantum}, and photonic circuits~\cite{xanaduXanaduWelcome, psiquantumPsiQuantumBuilding} are in a race to develop a commercial quantum computer, yet all face issues with qubit stability and accessibility. 

Motivated by the highly successful history of the classical cloud industry, a multi-node and distributed quantum system has been proposed~\cite{ang2022architectures}. This system aims to scale superconducting processors out by connecting multiple smaller quantum machines via M2O transduced optical links, facilitating the execution of larger circuits, offering a promising solution towards constructing large-scale distributed systems without the challenges associated with large monolithic devices. In resource-constrained quantum systems, each quantum worker is equipped with a limited number of qubits. Consequently, large circuits requiring more qubits than are available on individual workers cannot be executed directly. 
A quantum circuit can be represented as a Directed Acyclic Graph (DAG), where each vertex corresponds to inputs, outputs, or operations. These vertices are interconnected by directed edges that symbolize the flow of qubits. A large circuit, depicted by a DAG, can be pre-processed into smaller subcircuits, or subgraphs, allowing them to fit within the capacity of smaller quantum workers. Subsequently, the results from these subcircuits can be post-processed to reconstruct the algorithmic outcomes of the original circuit. There are two types of quantum circuit cuttings, wire cut~\cite{peng2020simulating,Tang_2021, brandhofer2023optimal, chen2023online, lowe2023fast} and gate cut~\cite{ufrecht2023cutting, bechtold2023investigating, mitarai2021constructing, fujii2022deep}. A wire cut refers to cutting along the 'wires' of a quantum circuit. Wires in a quantum circuit diagram represent the qubits over time, so a wire cut effectively separates qubits used in different parts of the computation. A gate cut focuses on cutting specific quantum gates that connect different parts of a circuit. 

These cutting methods, however, often encounter long search times when identifying optimal cutting points in quantum circuits. For instance, CutQC~\cite{Tang_2021}, a comprehensive pipeline for circuit cutting and reconstruction, can take from 100 to 1000 seconds to process large circuits containing 50 or more qubits. This is due to the fact that CutQC enumerates all the potential cutting options and attempts to find an optimal cutting point that minimizes the number of cuts.  
Moreover, the existing literature generally overlooks constraints such as qubit count on specific quantum machines. For instance, authors in~\cite{brandhofer2023optimal} aim to maximize sparsity by cutting circuits into the smallest possible subcircuits. However, within a specific quantum system, it is more advantageous for a circuit to fully utilize the available quantum workers and maximize qubit usage, rather than simply minimizing circuit size.

Leveraging community-based clustering, we propose a bottom-up approach, \sol, that transforms a DAG into a gate-based weighted graph. This graph is then clustered into communities to maximize modularity. \sol~iteratively merges these communities while considering quantum worker configurations, such as the varying number of available qubits. Ultimately, subcircuits are strategically scheduled across the workers in the distributed quantum system, aiming to maximize their qubit utilization rate. \sol~has been extensively evaluated under both single-node and distributed environments considering fidelity, runtime, and resource utilization.  
We summarize the key contributions below.

\begin{itemize}

\item We propose \sol~that transforms a quantum circuit into a gate-only-based graph representation. It employs a community-based algorithm designed to efficiently segment large circuits into smaller subcircuits while adhering to qubit constraints.

\item \sol~minimizes the number of cuts by merging communities and dynamically adjusting to the resource constraints of quantum workers. Its scheduling algorithm is tailored to distribute subcircuits in a way that maximizes resource utilization across each worker.

\item When compared to the Circuit Knitting Toolbox~\cite{circuit-knitting-toolbox}, \sol~demonstrates significant performance enhancements, reducing the time cost by factors ranging from 3 to 2000. Additionally, \sol~improves resource utilization rates by up to 3.88 times per individual worker and achieves a system-wide improvement of 2.86 times.

\end{itemize}

%% file: relatedworks.tex
\section{Related Works}

Manufacturing a single monolithic quantum device presents significantly greater challenges compared to building smaller machines with fewer qubits. More recently, researchers have spent efforts to explore multi-node and distributed quantum systems that connect multiple smaller quantum machines. Utilizing such distributed systems requires quantum circuit cutting, a technique that splits a large quantum circuit into smaller sub-circuits. 
This technique typically involves two types of cuts: wire cuts and gate cuts.

A gate cut aims to represent non-local gates using local operations, eliminating the need for concurrent execution of multiple qubits associated with the non-local gate. This is often used to resolve gate dependency which represents the requirement of qubits involved in the same multi-qubit gate to be connected. The concept of quasiprobability decomposition (QPD) and the analysis of sampling overhead for multi-qubit gates was initially introduced by Mitarai et al. in \cite{mitarai2021constructing,Mitarai_2021overhead}. Subsequent research efforts by Schmitt et al. \cite{schmitt2024cutting} and Piveteau et al. \cite{piveteau2023circuit} have focused on optimizing sampling overhead in this context.

A wire cut refers to cutting along the `wires' of a quantum circuit. It effectively separates qubits used in different segments of the computation, allows for separate executions, and utilizes classical computation to reconstruct the full probability distribution. The theory behind wire cutting was first proposed by Peng et al. \cite{peng2020simulating} with tensor network representations. Later, Tang et al. introduced CutQC, an end-to-end hybrid approach to wire cutting, as described in \cite{Tang_2021}. Subsequently, this approach was integrated into the Qiskit extension Circuit Knitting Toolbox. Several recent studies \cite{brenner2023optimal, chen2023online, lowe2023fast,harada2023doubly,pednault2023alternative} have explored various wire cutting methodologies aimed at achieving optimal sampling overhead under different scenarios. For instance, Lowe et al. \cite{lowe2023fast} introduced a modified wire cutting technique that leverages randomly sampled measurement and state preparation to reduce sampling overhead. Brenner et al. \cite{brenner2023optimal} investigated the optimal sampling overhead in scenarios where classical communication is enabled. Moreover, Harada et al. \cite{harada2023doubly} and Pednault et al. \cite{pednault2023alternative} demonstrated that optimal sampling overhead can be attained without the need for ancilla qubits.

In practical implementations, determining the optimal cutting points of an input quantum circuit is crucial, particularly when dealing with large-scale circuits where manual identification of such points becomes impractical. 
As part of the CutQC pipeline, an automatic cut searcher has been proposed \cite{Tang_2021}, utilizing a Mixed-Integer Programming (MIP) model where all elements, including edges and vertices, are represented as binary variables. CutQC aims to minimize the number of cuts with a mathematical solver that finds solutions adhering to hardware constraints. Its the search algorithm guarantees a globally optimal solution by exploring the entire solution space. Similarly, Brandhofer et al. \cite{brandhofer2023optimal} proposed a Satisfiability Modulo Theories (SMT) model employing a similar mechanism to find optimal partitioning, considering both gate cut and wire cut. However, a major drawback of these models lies in their factorial complexity, which is directly related to the number of edges in the DAG representation. Consequently, for deeper circuits with a larger number of edges, the search time increases significantly, posing a substantial limitation.

%% file: background.tex
\section{Background and Motivation}
\label{back}

\begin{figure*}[htbp]
  \centering
    \includegraphics[width=0.9\linewidth]{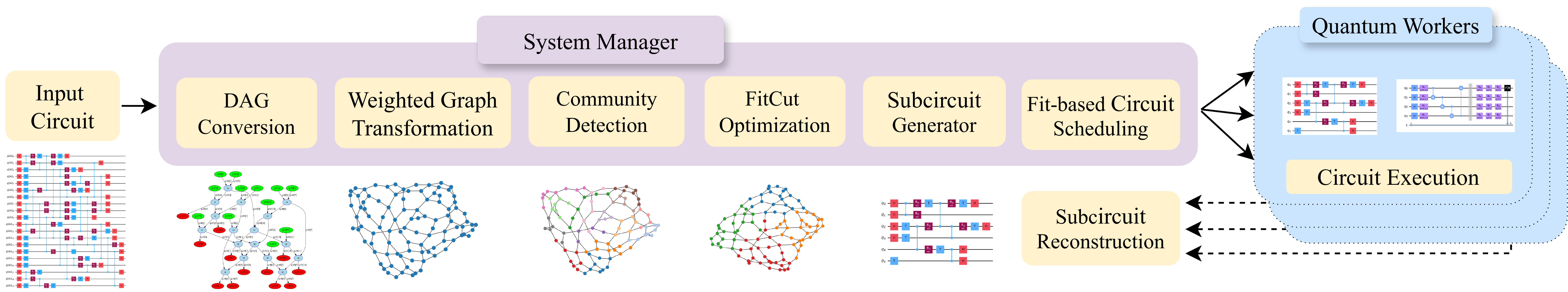}
    \caption{System Overview}
    \label{fig:Overview}
\end{figure*}

A quantum circuit, represented by a $DAG$, includes directed edges with vertices that signify both qubits and operations. Traditional graph-based methods such as the Kernighan-Lin Algorithm (KL), K-cuts, and Louvain communities are adept at dividing large graphs into smaller subgraphs. However, these methods are generally unsuitable for cutting quantum circuits since they could disrupt the operational flow within the $DAG$. Additionally, these algorithms have specific optimization objectives and constraints that may not align with the needs of quantum circuit partitioning. For instance, the KL Algorithm is designed to minimize edge cuts but is restricted to creating only two segments. Moreover, cutting a quantum circuit must also account for resource constraints like qubit counts, further complicating the application of these traditional methods.


Inspired by \cite{Blondel_2008}, we employ a modularity-based community detection algorithm as a preprocessing step to form initial partitions. Modularity is a heuristic metric in our setting, measuring the extent to which a graph can be segmented into distinct communities. They are characterized by densely connected vertices within each community and fewer connections between different communities. This approach helps in assessing the quality of initial partitions by analyzing their community structure, thereby facilitating the identification of significant divisions.


Given a weighted and undirected graph with a partition \([C_1,C_2,...,C_n]\) where each vertex is assigned to one and only one of the detected communities. Modularity is defined as:
\begin{equation}
Q(P) = \frac{1}{2m} \sum_{C}\sum_{i}\sum_{j} \left( A_{ij} - \frac{k_i k_j}{2m} \right) 
\label{eq1}
\end{equation}
\begin{itemize}
    \item vertices \( i \) and \( j \) are the vertices within community \(C\);
    \item \( A_{ij} \) is the weight of the edge between vertices \( i \) and \( j \);
    \item \( k_i \) and \( k_j \) are the degrees of vertices \( i \) and \( j \) respectively;
    \item \( m \) is the total weight of all edges in the graph.

\end{itemize}

Based on Equation~\ref{eq1}, the modularity of each individual community $C_i$ can be calculated as:

\begin{equation}
Q_{C_i} = \frac{\sum_{in}}{2m} - \left( \frac{\sum_{tot}}{2m} \right)^2
\label{eq2}
\end{equation}
\begin{itemize}
        \item \( \sum_{in} \) is the sum of edge weights between vertices within the community \( c \) (each edge is considered twice);
        \item \( \sum_{tot} \) is the sum of all edge weights for vertices within the community (including edges that link to other communities).
    \end{itemize}

The algorithm's objective is to maximize the modularity so that the communities are divided as sparsely as possible. It is a two-phased algorithm that includes, greedy optimization and agglomeration until the modularity score reaches a plateau, indicating convergence to a stable partition of the network.
A generalized workflow can be summarized as 3 steps:
(1) \textbf{Initialization}: each vertex is assigned to its own community.
(2) \textbf{Greedy optimization}: The vertices will be randomly shuffled to decide the order of iteration. for each vertex $i$, the change in modularity is calculated for removing $i$ from its own community and moving it into the community of each neighbor $j$ of $i$. Once the change in modularity $\displaystyle \Delta Q$ has been computed for all communities $\{c_{j}\}$ that vertex $i$ is connected to, vertex $i$ is placed into the community that resulted in the greatest modularity increase. If no increase is possible, vertex $i$ remains in its original community. We repeat this process sequentially to all vertices until no increase in modularity occurs.
(3) \textbf{Agglomeration}: Each community $c_j$ is reduced to a single vertex. Edges connecting vertices from $c_j$ to other communities and likewise reduced to a single weighted edge.


In the context of maximized modularity, each community is structured to have a minimized number of vertices. While this approach is effective for graph partitioning, it can lead to underutilized resources in quantum circuit cutting scenarios. For instance, on a quantum device equipped with 10 stable qubits, without considering noises, it is more resource-efficient to execute two 9-qubit circuits rather than six 4-qubit circuits, as the latter significantly under-utilizes available qubits.


To address this issue, during each agglomeration phase of our algorithm, we implement a predefined constraint, such as a qubit limit, to determine when to stop the iteration rather than solely focusing on maximizing modularity. This design allows our algorithm to be adaptable to a cluster of quantum machines with varying capacities. Our goal is to maximize the utilization rate of quantum resources effectively, without compromising the efficiency of circuit cuts.

%% file: solution-design.tex
\section{\sol~Solution Design}


As discussed in Section~\ref{back}, the directional edges and dual-typed vertices (qubits and gates) in a $DAG$ make it unsuitable for community-based algorithms. Therefore, we introduce \sol~ with a comprehensive suite of algorithms designed to convert a $DAG$ into a graph, partition a large quantum circuit into smaller subcircuits while considering qubit limitations and resource utilization, and efficiently schedule these subcircuits across a cluster of quantum machines.

Figure~\ref{fig:Overview} presents an overview of the \sol~system. Initially, a large circuit is fed into the system manager, which converts it into its $DAG$ representation. This $DAG$ is then passed to our Graph Transformation module, where it is converted into an undirected graph with two-qubit gates represented as vertices. The Community Detection is subsequently applied to this graph to maximize modularity within the constraints of qubit limits. Meanwhile, the FitCut Optimization module merges communities to optimize resource utilization based on specific qubit capacity on quantum workers. The merged communities are then transformed into subcircuits, which the FitCut Scheduler assigns to the appropriate workers. After the execution, workers send the results back to the manager, who reconstructs them to produce the final outcome. In the following subsections, we will detail the key modules and algorithms.

\subsection{Constrained $DAG$-Graph Weighted Transformation}

\begin{figure}[htbp]
  \centering
    \includegraphics[width=0.9\linewidth]{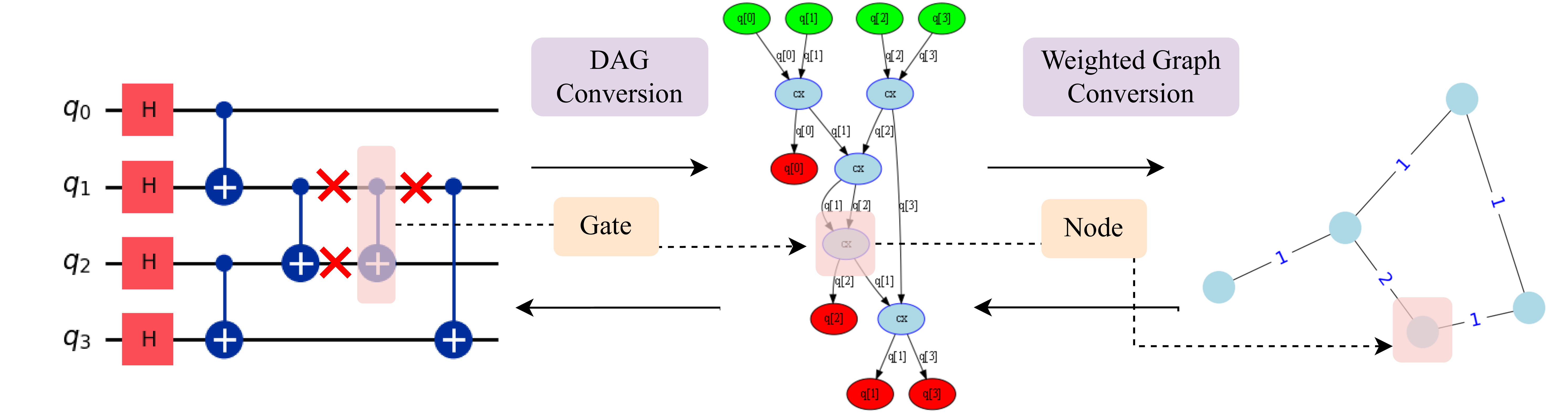}
    \caption{An Example of circuit transformation}
    \label{fig:Circuit example}
\end{figure}

A $DAG$-Graph transformation process is depicted in Figure \ref{fig:Circuit example}, which illustrates the transformation of gates into vertices and the computation of edge weights. The pink shaded CNOT gate within the quantum circuit is interconnected with other CNOT gates through three wires highlighted with red X marks. This connection is also shown in $DAG$. This results in the transformed vertex having a degree of 3. Additionally, when two edges connect to another identical CNOT gate, they combine into a weighted edge with a weight of 2.

Given a $DAG$ representation, we first remove all one qubit gates and convert it to a weighted graph $G_w(V,E)$ with the following variables and constraints:
\[V = \{v_{1}, ... ,v_{n}\} \]
\[ E = \{(v_i,v_j):v_i,v_j\in V\}\]
\[1\leq K_i \leq 4\]
\[0\leq A_{i,j} \leq 2\]  
\begin{itemize}
    \item The vertices V represent two-qubit quantum gates in the circuit.
    \item  The edges E represents the circuit wires that connect two-qubit gates.
    \item Having an edge between two vertices indicates that these two gates share qubits. The weight of edge $E_w$ is the number of qubits two vertices shared with a max number of 2.  
    \item $K_i$ represents the degree of vertices.
    \item \(A\) is the adjacency matrix and \( A_{ij} \) is the weight of the edge \( v_i, v_j \). 
\end{itemize}

\begin{figure}[htbp]
  \centering
    \includegraphics[width=0.65\linewidth]{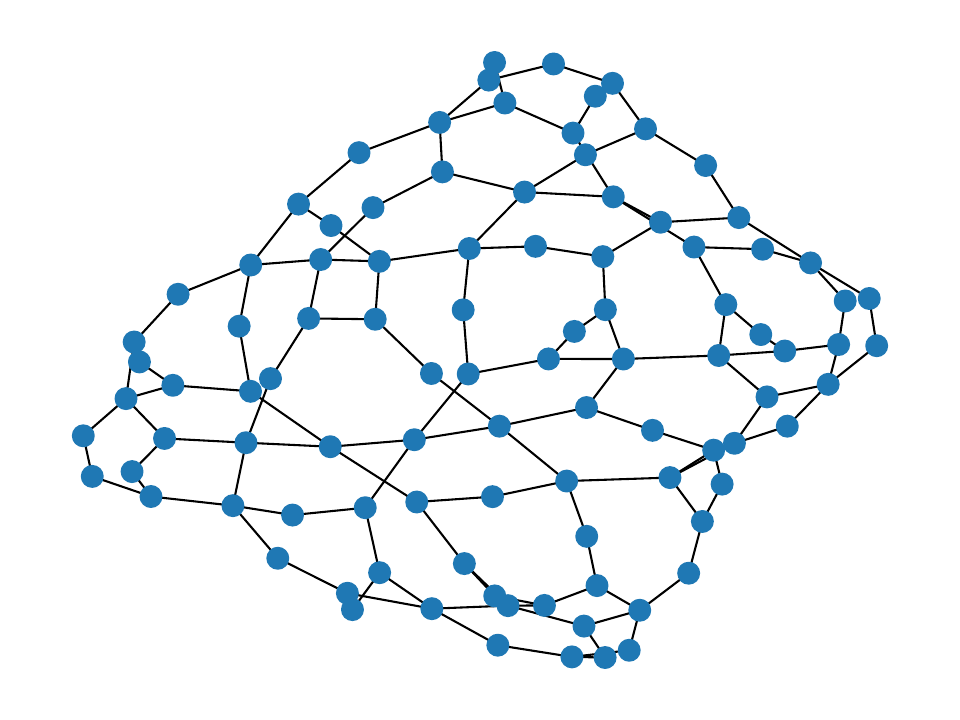}
    \caption{The Graph Representation of $7\times8$ Supremacy Circuit}
    \label{fig:supermacy_example}
\end{figure}

During the partition search in our system, edges within the graph are classified into two categories: Within-Subcircuit (WS) edges and Between-Subcircuit (BS) edges. The total weight of the BS edges directly correlates with the number of cuts required, which is a critical measure in our problem context. Our primary objective is to maximize the weight of the WS edges while minimizing the weight of the BS edges, all within the given constraints. The graph depicted in Figure \ref{fig:supermacy_example} illustrates a 7x8 supremacy circuit after undergoing the specified transformation. This graph effectively captures the connections between gates within the circuit, providing a clear visual representation of the circuit’s internal structure.

\subsection{Constrained Community Detection}

Based on modularity, we develop a constrained community detection method. At the end of each agglomeration phase, we calculate the number of qubits in each community $C_i$ with the following constraints:
\begin{equation}
q_{C_i} = 2 \times |C_i| - \sum_{v_i,v_j \in C_i} A_{ij} 
\label{eq:subcircuit_qubits}
\end{equation}

\begin{equation}
\label{eq:community_constraint}
q_{C_i} \leq \frac{1}{2}max(QC_{W}) 
\end{equation}
where community \(C_i\) is a set of vertices, and \(A_{ij}\) represents the weight of the edge that both endpoints are within the community. Constraint \eqref{eq:community_constraint} refers to the number of qubits associated with each community that cannot exceed half of the qubit capacity of the largest worker.


The community detection iteration terminates when the constraint is violated. This limitation is critical because merging two communities that each exceed half the worker's capacity would likely violate the qubit constraints, rendering further improvements in cut reduction and utilization rate unfeasible. Consequently, there is a fundamental trade-off between the flexibility required for subsequent phases and the complexity of the search process. Our goal is to finely tune the granularity of the resulting communities to strike a balance between these factors. 

\begin{figure}[ht]
\begin{subfigure}[b]{0.49\linewidth}
\centering
\includegraphics[width=\textwidth]{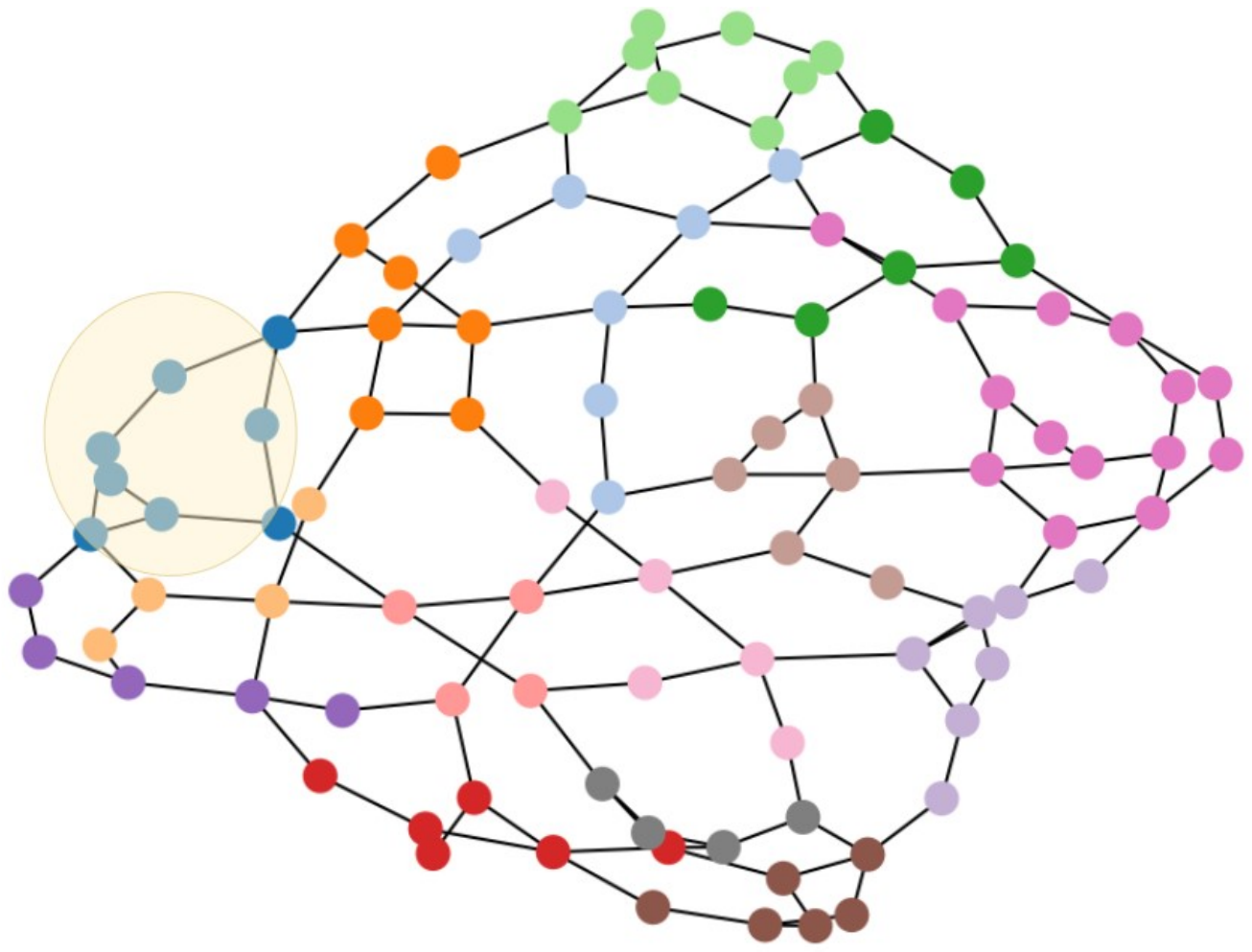}
\caption{}
\label{fig:colored_community}
\end{subfigure}
\begin{subfigure}[b]{0.49\linewidth}
\centering
\includegraphics[width=\textwidth]{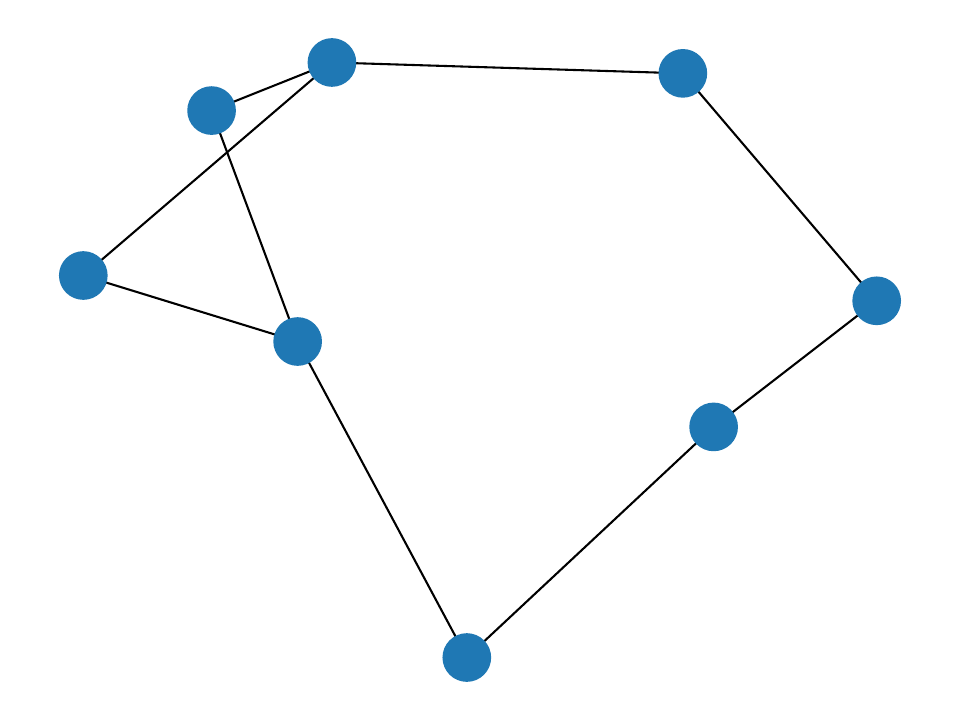}
\caption{}
\label{fig:subgraph_example}
\end{subfigure}
\caption{(a) Constrained Community Detection and (b) Subgraph of one Community}
\end{figure}

The community detection results for the $7\times8$ supremacy circuit example are shown in Figure \ref{fig:colored_community}, where vertices within the same community share identical color codes. The highlighted community in Figure \ref{fig:colored_community} is further detailed in Figure \ref{fig:subgraph_example}. This example demonstrates the internal structure of a super vertex in the subsequent step.

\subsection{\sol~Problem Formulation}


Our goal is to minimize the sampling overhead and provide an effective subcircuits scheduling, which maximizes the system resource utilization rate while adhering to resource constraints. Therefore, the workers and their qubit capacities denoted as $W:\{(W_1,QC_{W_1}),(W_2,QC_{W_2}),\dots,(W_m,QC_{W_m})\}$ are taken into consideration in next phase \sol~optimization. 

The communities identified in the detection step serve as the baseline partition. Each individual community now represents the smallest unit to be merged together to form a larger subcircuit while satisfying the resource constraints. As the Within-Subcircuit
(WS) edge weights do not influence the connectivity between subcircuits, all edges within individual communities are omitted in this step. Conversely, the Between-Subcircuit (BS) edge weights are crucial for determining the number of cuts. Therefore, we agglomerate each community into a single super vertex, with its weight equal to the number of qubits associated that community. The resulting edges between super vertices are potential BS edges in our final partition. The merged graph $G_m(SV,SE)$ is characterized by the following variables:

\[SV = \{sv_{1}, ... ,sv_{n}\} \]
\[ SE = \{(sv_i,sv_j):sv_i,sv_j\in SV\}\] 
\[SVW = \{wt_{sv_1}, ... ,wt_{sv_n}\} \]
\[SEW = \{wt_{(sv_i,sv_j)}:(sv_i,sv_j)\in SE\} \]
\[P = \{(P_1,QC_{P_1}),(P_2,QC_{P_2}),\dots,(P_n,QC_{P_n})\}, P_i = [sv_i] \] 
\[QC_{P_i} = \sum_{sv_i \in P_i} {wt_{sv_i}} - \sum_{sv_i,sv_j \in P_i} wt_{(sv_i,sv_j)}\]
\[QC_{P_i} <= max(QC_{W_i})\]
\begin{itemize}
    \item $SV$ represents the set of super vertices.
    \item $SE$ represents the weighted edges. \item $SVW$,$SEW$ are the corresponding weights of vertices and edges, respectively. The vertex weight $wt_{sv_1}$ signifies the number of qubits within each community, while the edge weight $wt_{(sv_i,sv_j)}$ is equivalent to the number of shared qubits between two super vertices.
    \item $P$ represents a partition of $G_m$ such that each super vertex ($sv$) belongs to exactly one partition $P_i$. Each $P_i$ corresponds to a set of vertices representing a subcircuit with $QC_{P_1}$ qubit numebr. 
    \item $QC_{P_i}$ denotes the number of qubits for the subcircuit, calculated as the sum of qubits of each super vertex minus the sum of Within-Partition edge weights. The imposed constraint is that $QC_{V_i}$ does not surpass the maximum qubit capacity of our system.

\end{itemize}

\begin{figure}[htbp]
  \centering
  \includegraphics[width=0.70\linewidth]{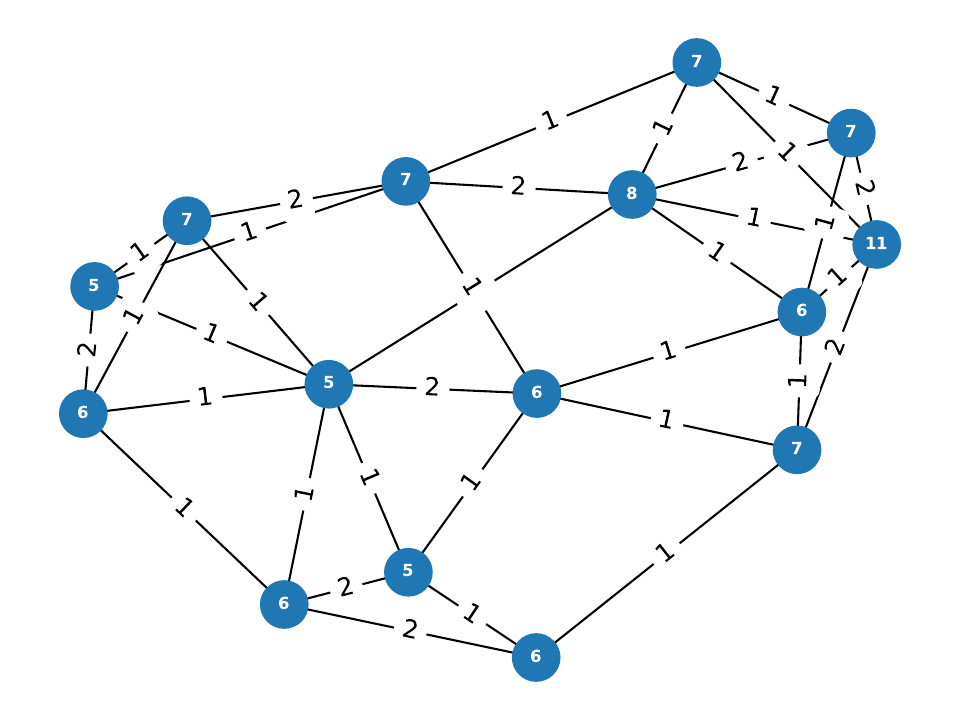}
    \caption{Merged Graph with SVWs and SEWs}
    \label{fig:ret_example}
\end{figure}

The merged graph from detected communities of  the supremacy circuit example \ref{fig:colored_community} is shown in figure \ref{fig:ret_example}. Each vertex in \ref{fig:ret_example} is a super vertex $sv_i$ labeled with corresponding $wt_{sv_i}$. Each edge$(sv_i,sv_j)$ is labeled with $wt_{(sv_i,sv_j)}$.

Iteratively, \sol~evaluates the possibility of relocating each vertex from its current community to a neighboring community. Each relocation will result in an updated partition $P_{update}$.
\begin{align*}
    & P_{update}: \{(P_1,QC_{P_1}),(P_2,QC_{P_2}),\dots,(P_n,QC_{P_n})\} \\
    & W: \{(W_1,QC_{W_1}),(W_2,QC_{W_2}),\dots,(W_m,QC_{W_m})\}
\end{align*}
With updated partition $P_{update}$ and system configuration $W$, each worker $W_i$ will be assigned with a list of subcircuits $\{P_i,...\}$ by calling circuit scheduling algorithm~\ref{alg:Circuit Scheduling}. 
Based on the resulting scheduling, our optimization process adheres to a two-tiered objective function calculated by algorithm \ref{alg:OBJ}. Primarily, \sol~aim to minimize the number of cuts in the circuit, which is defined as:
\begin{equation}
    Obj_{\text{nc}} = \sum_{P_i \in P}(QC_{P_i}) - QC_{input}
\end{equation}
This definition is valid because each cut introduces a new initialization qubit in a subcircuit. Consequently, the total number of extra qubits corresponds to the number of cuts.
In the event of a tie, the second objective is to prioritize maximizing the system resource utilization rate, which is defined as minimizing the idling resources during the execution of subcircuits. For each worker, $Obj_{ru}(W_i)$ calculate the number of idling qubits associated with its assigned subcircuit and sum them up across all workers.
\begin{equation}
    Obj_{ru}(W_i) = \sum_{P_i \in W_i} (QC_{W_i} - QC_{P_i})
\end{equation}

\begin{figure}[htbp]
  \centering
    \includegraphics[width=0.70\linewidth]{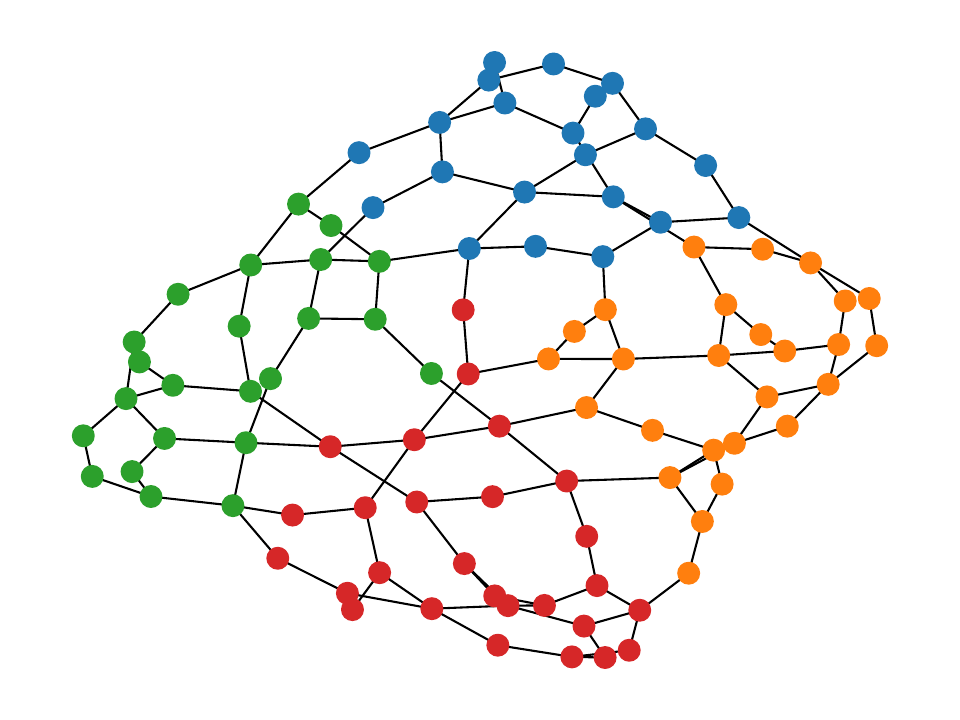}
    \caption{Final Partition by \sol}
    \label{fig:fit-cut}
\end{figure}

The iteration concludes when no actions are taken during one complete round of iteration over all super vertices. The entire process is detailed in Algorithm \sol~Optimization \ref{alg:FITcut}. The final partition result given by FitCut method is shown in figure \ref{fig:fit-cut}. With the worker capacity being 20-qubit for a 56-qubit supermacy circuit, the result subcircuits are 20-qubit, 19-qubit, 17-qubit and 16-qubit quantum circuit.

In summary, \sol~aims to partition a weighted graph, that represents a large quantum circuit, into smaller subcircuits that align with our worker configurations. It first transforms an input circuit into a weighted graph as dipicted in \ref{fig:Circuit example} and applied constrained community detection to obtain a baseline community structure as in \ref{fig:colored_community}. Subsequently, a merged graph $G_m$ is derived from the previous result and used as input for \sol~Optimization Algorithm \ref{alg:FITcut}, which takes the system configuration into account and aims to minimize the sampling overhead of wire cutting while maximizing the system resource utilization rate. During the optimization process, two algorithms Weighted Closest-First Circuit Scheduling~\ref{alg:Circuit Scheduling} and two-tiered objective function~\ref{alg:OBJ} are employed concurrently. The algorithms will be discussed in the following subsection.

%% file: algorithm.tex
\subsection{\sol~ Algorithm Design}
\begin{table}[ht]
	\centering
	\caption{Notation Table}
	\scalebox{0.65}{
		\begin{tabular}{ | m{5cm} | m{7cm}  |  }				
		    \hline
            $W=[(W_1,QC_{W_1}),\dots,(W_m,QC_{W_m})]$ & list of all workers, worker $W_m$ has $QC_{W_m}$ qubit \\
            \hline
            $P=[(P_1,QC_{P_1}),\dots,(P_n,QC_{P_n})]$ & list of all partition, each partition $P_n$ has $QC_{P_n}$ qubit \\
            \hline
            \centering $A$ & a dictionary for subcircuits scheduling. Key is $W_m$, and Value is  a list of $P_n$ assigned to $W_m$ \\
            \hline
            \centering $FC (W)$ & Find the worker with the closest minimum capacity $QC_{W_{closest}}$ that exceeds the subcircuits with $QC_{V_i}$ \\
            \hline
            \centering $WP$ & a list of all possible workers with the greater and equal capacity than $W_{closest}$\\
            \hline
            \centering $Max, Min$ & maximum and minimum number of subcircuits each worker should be assigned  \\
            \hline
            \centering $MO (W_i, W_j)$ & Moving function that for moving scheduled subcircuits from worker $W_i$ to worker $W_j$  \\
            \hline
            \centering \objOne, ~~\objTwo & the objective function for qubit constraint and the objective function for resource utilization \\
            \hline
            \centering $BO_1, BO_2$ & the current best value of two objective value \\
            \hline
            
		\end{tabular}	
	}
	\label{table:notation}
\end{table}  

\begin{algorithm}[!t]
\caption{Weighted Closest-First Circuit Scheduling, CS}
\begin{algorithmic}[1]
\STATE Input: 
\item[] $W=\{(W1,QC_{W_1},...,(W_m,QC_{W_m})\}$: workers
\item[] $P=\{(P_1,QC_{P_1}),\dots,(P_n,QC_{P_n})\}$: partition
\STATE Initialization: 
\item[] $A=\{W1:[\,],W2:[\,],...,Wm[\,]\}$: subcircuit scheduling
\FORALL{$P_i \in V$}
    \STATE $W_{closest} = FC(W, P_i)$
    \STATE $A[W_{closest}] \gets P_i$
\ENDFOR
\STATE $sort(A,\,reverse = True,\,key = QC_{W_i})$
\FOR{$W_i \in A$}
\STATE $WP = [\,]$
    \FORALL{$W_j \in A$}
        \IF{$QC_{W_j} >= QC_{W_i}$}
            \STATE $WP \gets W_j$
        \ENDIF
    \ENDFOR
    \STATE $Max = ceil(\sum_{W_p \in WP}|A[W_p]|\,/\, |WP|)$
    \STATE $Min = Max - 1$
    \STATE $sort(WP,\,key = QC_{W_p})$
    \FORALL{$W_p \in WP$}
            \WHILE{$|W_{i}| > Max$ and $W_P < Min$}
                \STATE $A \gets MO(W_i, W_p)$
            \ENDWHILE
            \WHILE{$|W_{i}| > Max$ and $W_P < Max$}
                \STATE $A \gets MO(W_i, W_p)$
            \ENDWHILE
    \ENDFOR
\ENDFOR
\RETURN $A$
\end{algorithmic}
\label{alg:Circuit Scheduling}
\end{algorithm}
In this subsection, we discuss the algorithms in \sol~ for circuit scheduling, updating the objective function and optimization. 

\subsubsection{Circuit Scheduling}

Algorithm~\ref{alg:Circuit Scheduling} illustrates the Weighted Closest-First Circuit Scheduling, aiming to optimize resource utilization on workers. The algorithm operates by assigning each super vertex$sv_i$(subcircuit), that is considered as partition $P_i$, to workers based on their qubit capacity, minimizing idling qubits and maximizing system utilization. Algorithm~\ref{alg:Circuit Scheduling} accepts two input lists: one containing the qubit numbers of the workers configured in our system, and the other containing the qubit numbers of the partition. It generates a dictionary that maps each worker to a list of scheduled subcircuits. The input data comprises two lists: $W$ representing the qubit numbers of configured workers, and $P$ containing the qubit numbers of the partition (Line 1). Next, the algorithm initializes a dictionary to store partition scheduling (Line 2). Initially, each partition is scheduled to the worker whose qubit capacity is closest to the partition's requirement (Lines 3-5). Next, it sorts all workers in $A$ from maximum to minimum by their respective qubit capacities ($QC_{W_i}$) (Line 6). For each worker containing scheduled partition (Lines 7-19), an empty list $WP$ is initialized to track possible workers (Line 8). Then, it iterates over each worker $W_j$ with a capacity greater than or equal to $QC_{W_i}$ in $WP$ (Lines 9-11). Maximization of system utility involves calculating the maximum and minimum number of partition each worker should handle, based on the total number of partition in $W_p$ and the number of workers in $WP$ (Lines 12-13)
Meanwhile, it sorts all workers in $WP$ from minimum to maximum by their respective qubit capacities ($QC_{W_p}$) (Line 14). The algorithm redistributes partition among workers to balance the workload (Lines 15-19), starting by identifying workers with minimal capacity that can accommodate a given partition (Lines 16-17). It update $A$ by function $MO$ that move one scheduled partition from worker $W_i$ to worker $W_p$ (Line 17). Subsequently, the algorithm moves jobs from the current worker to others in order to maximize the number of jobs assigned (Lines 18-19). Upon completion, the algorithm produces a dictionary indicating which job is assigned to each worker (Line 20).

\begin{algorithm}[!t]
\caption{Two-tiered Objective Function, OBJ}
\begin{algorithmic}[1]
\STATE Input: 
\item[] $A$: Subcircuits Scheduling
\item[] $QC_{input}$: number of qubits of input circuit
\STATE Initialization: \objOne~$ = 0$, \objTwo~$ = 0$
\STATE \objOne~$ = \sum_{V_i \in A}(QC_{P_i}) - QC_{input}$
\FOR{$W_i$ \textbf{in} $A$}
    \STATE $diff = \sum_{V_i \in A[W_i]} (QC_{W_i} - QC_{P_i})$
    \STATE \objTwo~$\mathrel{+}= diff$
\ENDFOR
\RETURN \objOne~$,$\objTwo~
\end{algorithmic}
\label{alg:OBJ}
\end{algorithm}

\subsubsection{Objective Function}
Algorithm~\ref{alg:OBJ} outlines our algorithm for calculating the objective function. It aims to evaluate two key objectives. The first objective aims to minimize the qubit constraint. The second objective aims to maximize resource utilization by minimizing the sum of the differences between each worker's capability and the number of qubit in the subcircuits assigned to that worker. The objective function serves as a guidepost for determining the necessity of proceeding to the next iteration. 

The input data for this algorithm consists of a dictionary named Circuit Scheduling $A$ and number of qubits of input circuit $QC_{input}$ (Line 1). Subsequently, the algorithm initializes two objective values, denoted as \objOne~ and \objTwo~ (Line 2). The first objective value (Lines 3) is computed by aggregating the number of qubits across all partition minus the input circuit's qubit count. The second objective value (Lines 4-6) is determined by summing the disparities between the number of qubit of each partition and the qubit capacities of their respective assigned workers. Finally, the algorithm returns these two objective values (Line 7).

\begin{algorithm}[!t]
\caption{\sol~ Optimization}
\begin{algorithmic}[1]
\STATE Input: 
\item[] $G_m=[C(SV,SE)]$: Graph 
\item[] $W=[(W_1,QC_{W_1}),\dots,(W_m,QC_{W_m})]$: System Configuration
\STATE Initialize: 
\item[] $P = \{[(P_1,QC_{P_1}),\dots,(P_n,QC_{P_n})], P_i = [sv_i]\}$ : partition
\item[] $C_{Nbr}=\{p_1:[P_i,\dots],\dots, p_i:[P_j,\dots]\}$ Neighboring Communities

\STATE Improvement is $True$
\WHILE{Improvement}
\STATE Improvement is $False$
\FOR{$P_i \in P$}
    \STATE $BO_1, BO_2 = Inf, Inf$
    \FOR{$p_i$ in $C_{Nbr}$}
        \STATE $P_{update} = copy(P)$
        \STATE  $P_{update}[v_i] = V_j$
        \STATE $P_{update}:\{(P_1,QC_{P_1}),\dots\} \gets QC_{P_i}$
        \IF{$max(QC_{P_i}) < max(QC_{W_i})$}
            \STATE $A_{update} \gets CS(W,P)$
            \STATE \objOne~,\objTwo~$\gets OBJ(A)$
            \IF{\objOne~$ < BO_1$}
                \STATE $P = P_{update}$
                \STATE $BO_1, BO_2 = $\objOne~,\objTwo~
                \STATE Improvement is $True$
            \ELSIF{\objOne~$ == BO_1$}
                \IF{\objTwo~$ < BO_2$}
                    \STATE $P = P_{update}$
                    \STATE $BO_2 = $\objTwo~
                    \STATE Improvement is $True$
                \ENDIF
            \ENDIF
        \ENDIF
    \ENDFOR
\ENDFOR
\ENDWHILE
\STATE $A \gets CS(P[P_i])$
\RETURN $P$, $A$
\end{algorithmic}
\label{alg:FITcut}
\end{algorithm}

\subsubsection{\sol~Optimization}
Algorithm~\ref{alg:FITcut} \sol~Optimization aims to identify optimal cutting locations given a weight graph representation of an input quantum circuit, with the objective of minimizing the number of cuts and efficiently distributing the workload among workers. The algorithm accepts the graph $G$ and system configuration $W$ as input (Line 1). During the initialization step, each vertex of the graph is assigned to its own partition, denoted by $P$. Additionally, the neighboring community $C_{Nbr}$ to which at least one vertex in the community is connected to $p_i$ is identified for each vertex $P_i$ (Line 2). This information is utilized as potential target communities for subsequent iterations.

The algorithm proceeds by iterating through the vertices, wherein it evaluates the objectives associated with the resulting list of subcircuits obtained by relocating each vertex to a neighboring community (Lines 4-23). The termination criteria is initially set in (Line 3). If no actions are taken during one complete iteration through all vertices, the algorithm will terminate.
For each vertex, it initializes both objective function values, \objOne~,\objTwo~, to infinity (Line 7). Next, a relocation from current partition to its neighboring communities were considered (Line 8 - 10). This relocation will form a new set of subcircuits $P_update$ with its corresponding number of qubits $QC_{V_i}$ using equation \ref{eq:subcircuit_qubits} (Line 11). To ensure hardware constraints are met, the algorithm checks whether the maximum number of qubits in any subcircuit is less than the maximum qubit capacity of the workers (Line 12).
Consequently, the updated scheduling $A_{Update}$ is obtained through circuit scheduling algorithm \ref{alg:Circuit Scheduling} (Line 13). Subsequently, we evaluation this relocation by calculating the objectives using algorithm \ref{alg:OBJ}. Using the two-tiered objective function, the algorithm updates the communities under two conditions: firstly, if the primary objective of reducing the number of cuts is achieved or maintained, and secondly, if the primary objective remains unchanged and the secondary objective is minimized. Consequently, the partition is updated accordingly and best objective value is recorded (Lines 15-23). The boolean variable "improvement" indicates whether any action has been taken during a complete round of iteration (Line 18,23). Ultimately, the algorithm scheduling subcircuit represents the $C[V_i]$ (Line 24),
returns the final solution $A$ and $C$ that consisting of multiple communities that have been optimized for workload balance (Line 25).

\subsection{Reconstruction}
\label{reconstruction}
After obtaining the measurement result of subcircuits following execution on quantum workers, we can combine it to reconstruct the probability distribution or Pauli observable of our interest for input circuit. Based on the theory provided by \cite{peng2020simulating}, in an ideal noise-free simulation, the results are expected to be perfectly reconstructible. Along with experimental findings in \cite{Tang_2021}, executing subcircuits on a NISQ machine and reconstructing them back to the input circuit leads to improved fidelity. This enhancement is attributed to the subcircuits being both smaller and shallower in comparison to the uncut circuit. 

%% file: evaluation.tex
\section{Evaluation}
In this section, we evaluate our system with different workloads and configurations. 

\subsection{Workload, Implementation and Experiment Settings}

\noindent {\bf Workload}: We evaluate \sol~ system with following circuits in different settings.
\textbf{Adder} circuit is a linear-depth ripple-carry quantum addition circuit that performs addition of quantum states \cite{cuccaro2004new}. It facilitates arithmetic operations with quantum parallelism and leveraging quantum entanglement. The adder quantum circuit requires an even number of qubits because it can only add two quantum states of the same width.
\textbf{Bernstein-Vazirani} (BV) circuit solves hidden binary strings encoded by secret functions using a single query to the black-box oracle, demonstrating quantum advantage of exponential speedup compared with classical methods \cite{bernstein1993quantum}.
\textbf{Hardware-efficient ansatz} (HWEA) is a parameterized quantum circuit architecture designed to efficiently represent quantum states on near-term quantum devices. It aims to minimize the number of gates and optimize gate connectivity to enhance performance while considering hardware constraints \cite{Kandala_2017}.
\textbf{Supremacy} circuit is a 2-D random circuit example from \cite{Boixo_2018} with dense probability output. It is used by Google to demonstrate a quantum advantage in \cite{Arute_2019}. We focused on near-square shapes, with two dimensions differing by up to 2 qubits (e.g., $7\times 8$), as they pose a more challenging task to find cuts.

\noindent {\bf Implementation and Platform}: We implement \sol~ system with the following software dependencies, Python 3.9, IBM Qiskit 1.02, Circuit Knitting Toolbox 0.6.0, Networkx 3.3, IBM ILOG CPLEX Optimization Studio 22.1.1.0
The platform we conduct our experiments is equipped with an AMD Ryzen 7 6800H processor running at 3.2 GHz.

\noindent {\bf Experiment Settings}: 
We compare \sol~ with the automatic cut searcher for wire cutting in Qiskit Circuit Knitting Toolbox (CKT)~\cite{circuit-knitting-toolbox}.
It operates by generating a Mixed-Integer Programming model based on the $DAG$ of the input circuit. Subsequently, it utilizes a mathematical solver (i.e., IBM ILOG CPLEX Optimization Studio) equipped with multiple optimization techniques. The primary objective of this cut searcher is to minimize the number of cuts.

To conduct experiments on runtime performance, We executed the task of cutting the identical input circuit to fit within the same qubits constraints on CKT and \sol. It is worth noting that CKT requires an additional input: the pre-defined number of subcircuits. Without prior knowledge regarding the optimal number of subcircuits for the final solution, the number of subcircuits is provided as a list of possible values starting from the lowest possible number of subcircuits which is defined as:
\begin{equation}
    \arg\min_{N_{\text{subcirc}}} QC_{\text{worker}}*N_{\text{subcirc}} \geq QC_{\text{input}} + N_{\text{subcirc}} - 1
\end{equation}
where $QC_{\text{worker}}$ is the qubits constraint of workers, $N_{\text{subcirc}}$ is the number of subcircuits, and $QC_{\text{input}}$ is the number of qubits of the input circuit. 
If no feasible solution under the predefined number of subcircuits, CKT increases the number of subcircuits by one and restarts the search process.

Furthermore, a runtime limit of 300 seconds has been enforced by Qiskit CKT for each number of subcircuits value. If the model has not completed the search space when the time limit is reached, it will return the best solution found within that timeframe. In complex cases, reaching the 300-second limit may result in the model returning no solution. This indicates that either there are no possible results within the given constraints, or the model was unable to find a solution within the allotted time.

\subsection{Fidelity}
As discussed in Section~\ref{reconstruction}, \sol~ employs the same subcircuit reconstruction procedure as previously described to obtain final results. Additionally, the transformation of a $DAG$ to a weighted graph and vice versa is reversible. Based on our experiments, we observe that \sol~ achieves the same fidelity as the Qiskit CKT. This demonstrates the effectiveness of \sol. Due to page limit and to avoid redundancy with prior works, we omit detailed fidelity comparisons.

\begin{figure*}[htbp]
\centering
    \begin{subfigure}[b]{0.23\textwidth}
        \includegraphics[width=\textwidth]{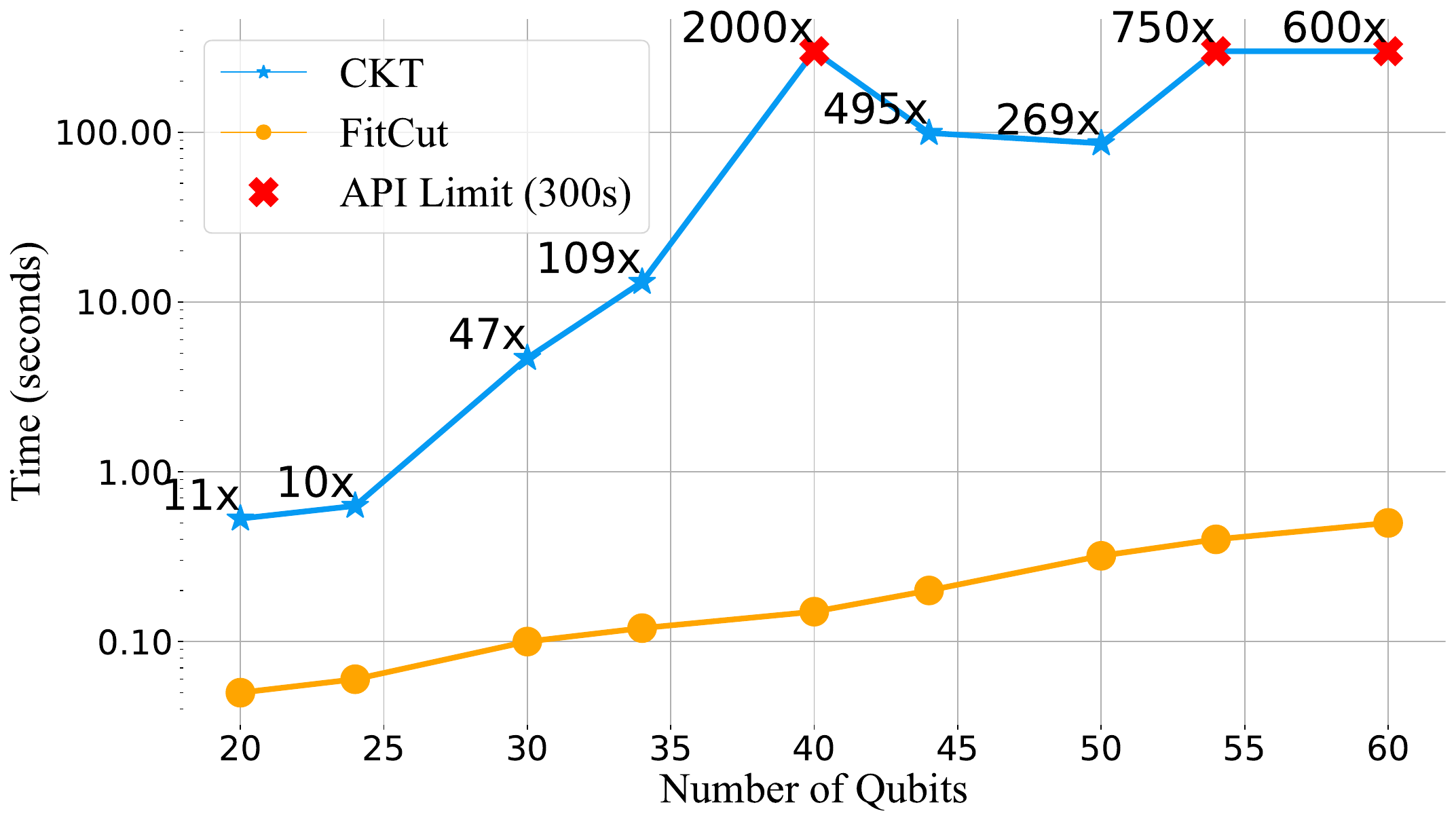}
        \caption{Adder}
        \label{fig:adder_run_15}
    \end{subfigure}
  \hfill
    \begin{subfigure}[b]{0.23\textwidth}
        \includegraphics[width=\textwidth]{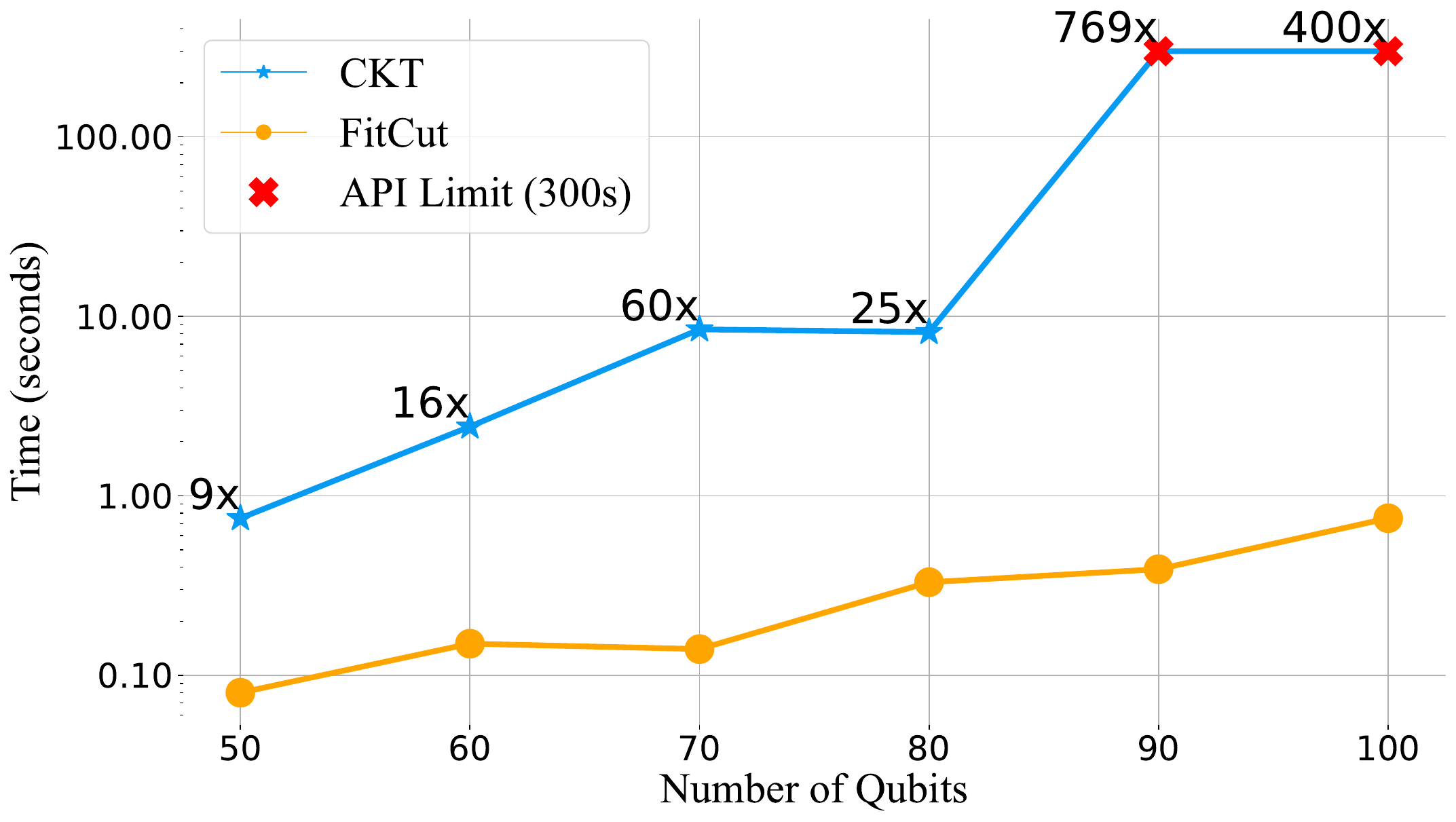}
        \caption{BV}
        \label{fig:bv_run_15}
    \end{subfigure}
  \hfill
    \begin{subfigure}[b]{0.23\textwidth}
        \includegraphics[width=\textwidth]{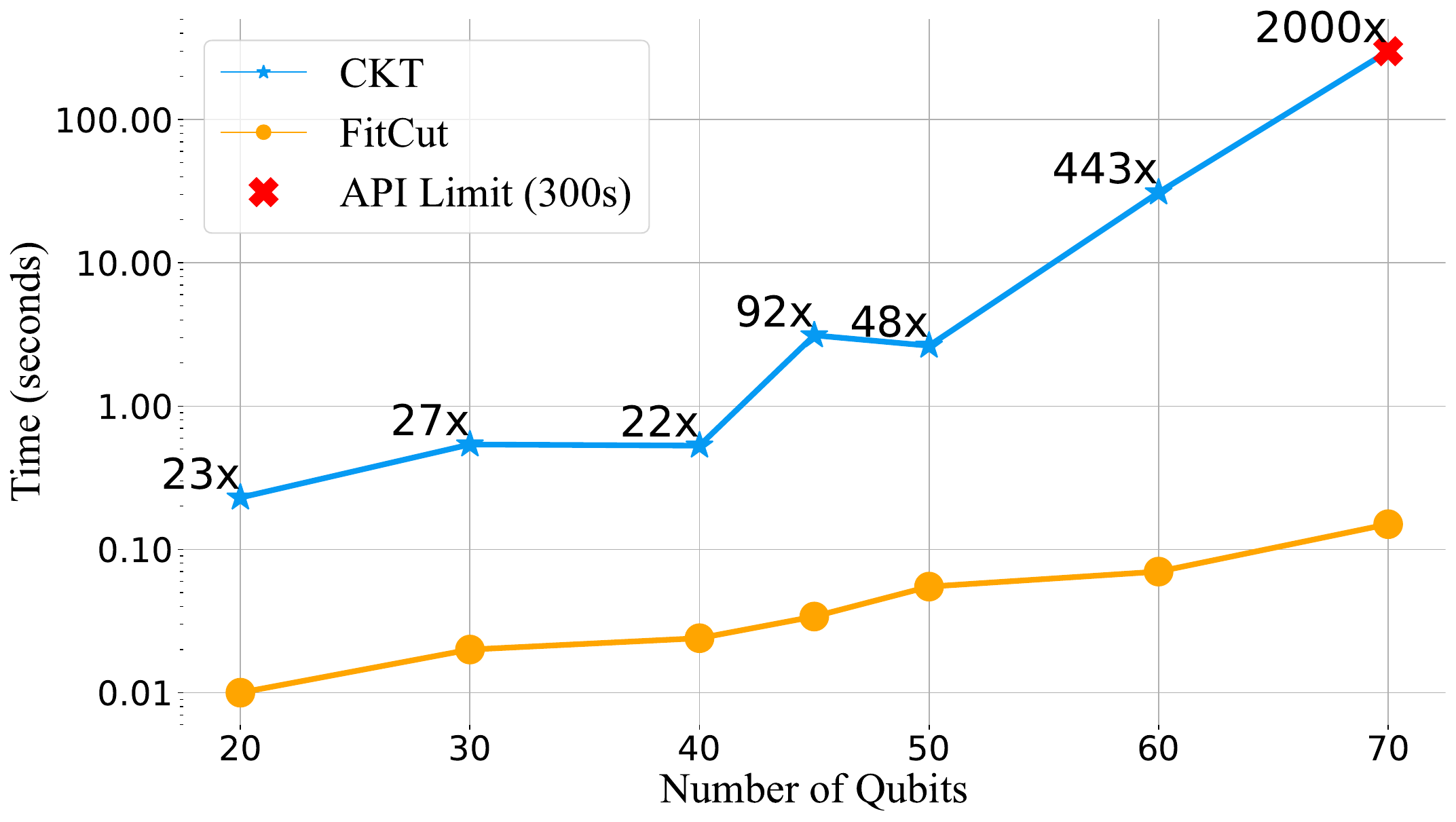}
        \caption{HWEA}
        \label{fig:hwea_run_15}
    \end{subfigure}
  \hfill
    \begin{subfigure}[b]{0.23\textwidth}
        \includegraphics[width=\textwidth]{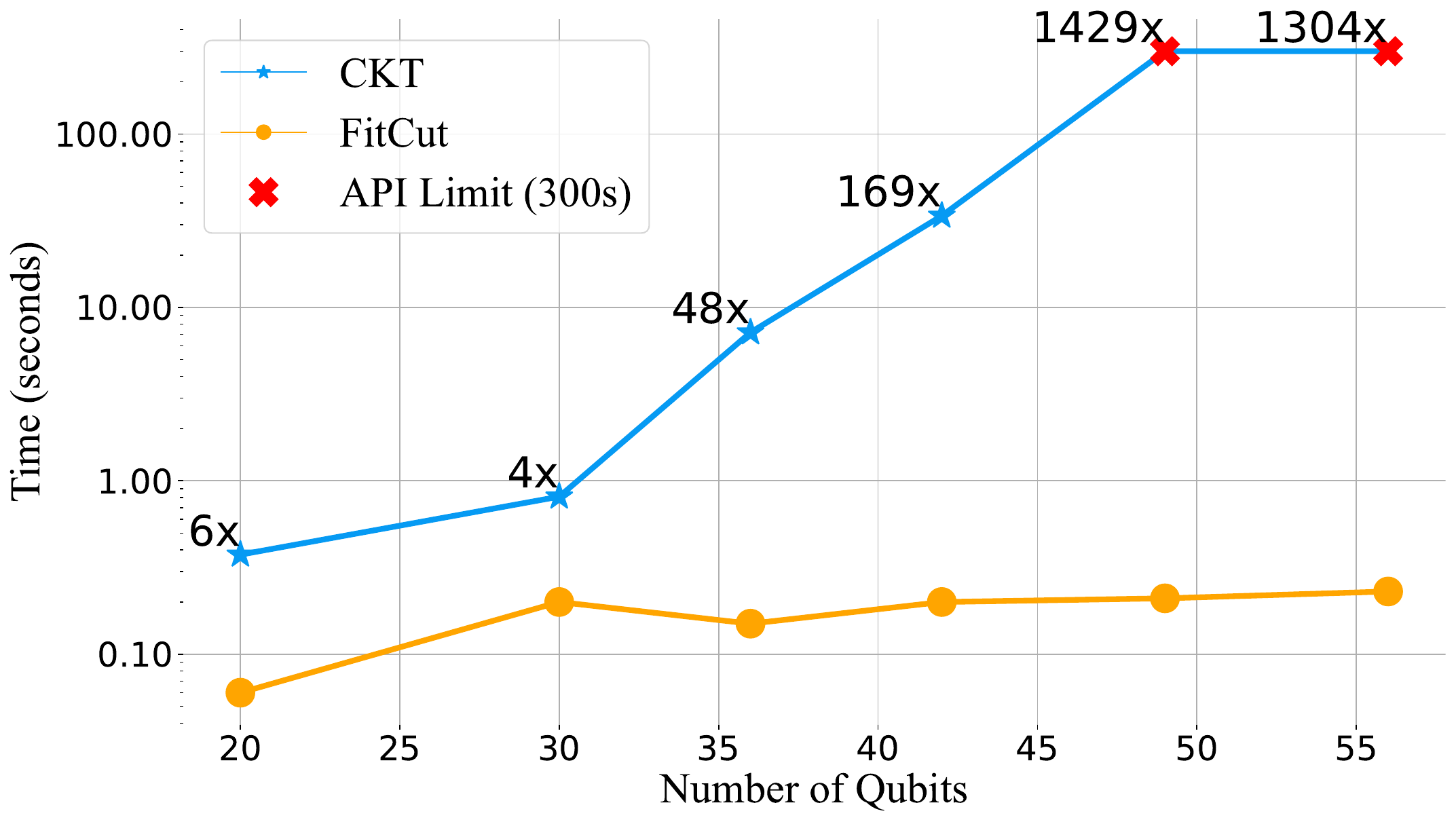}
        \caption{Supremacy}
        \label{fig:supremacy_run_15}
    \end{subfigure}

\caption{Runtime comparison in worker with 15-qubit}
\label{fig:worker15}
\end{figure*}

\begin{figure*}[htbp]
    \centering
        \begin{subfigure}[b]{0.23\textwidth}
        \includegraphics[width=\textwidth]{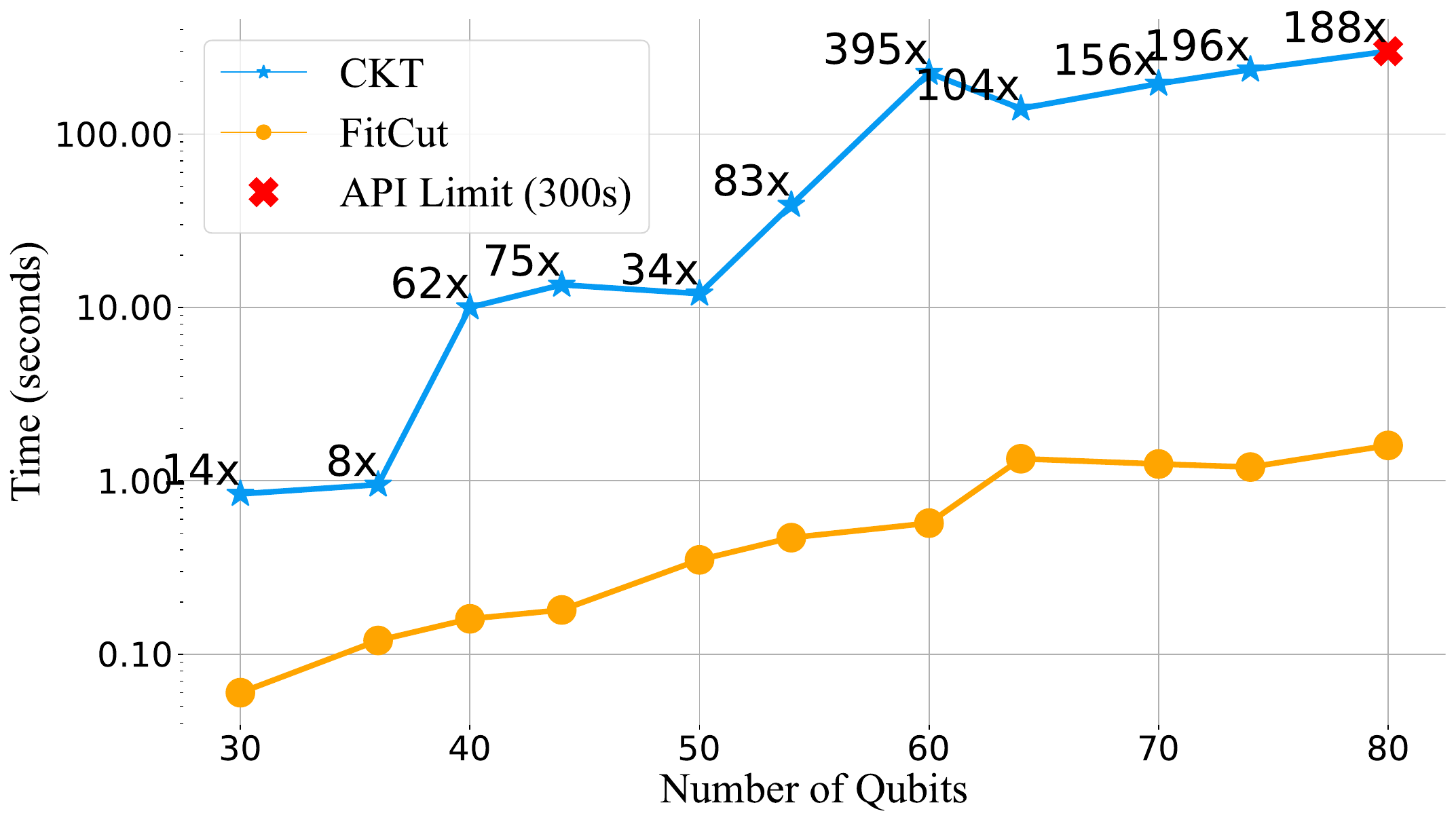}
        \caption{Adder}
    \label{fig:adder_run_20}
    \end{subfigure}
  \hfill
    \begin{subfigure}[b]{0.23\textwidth}
        \includegraphics[width=\textwidth]{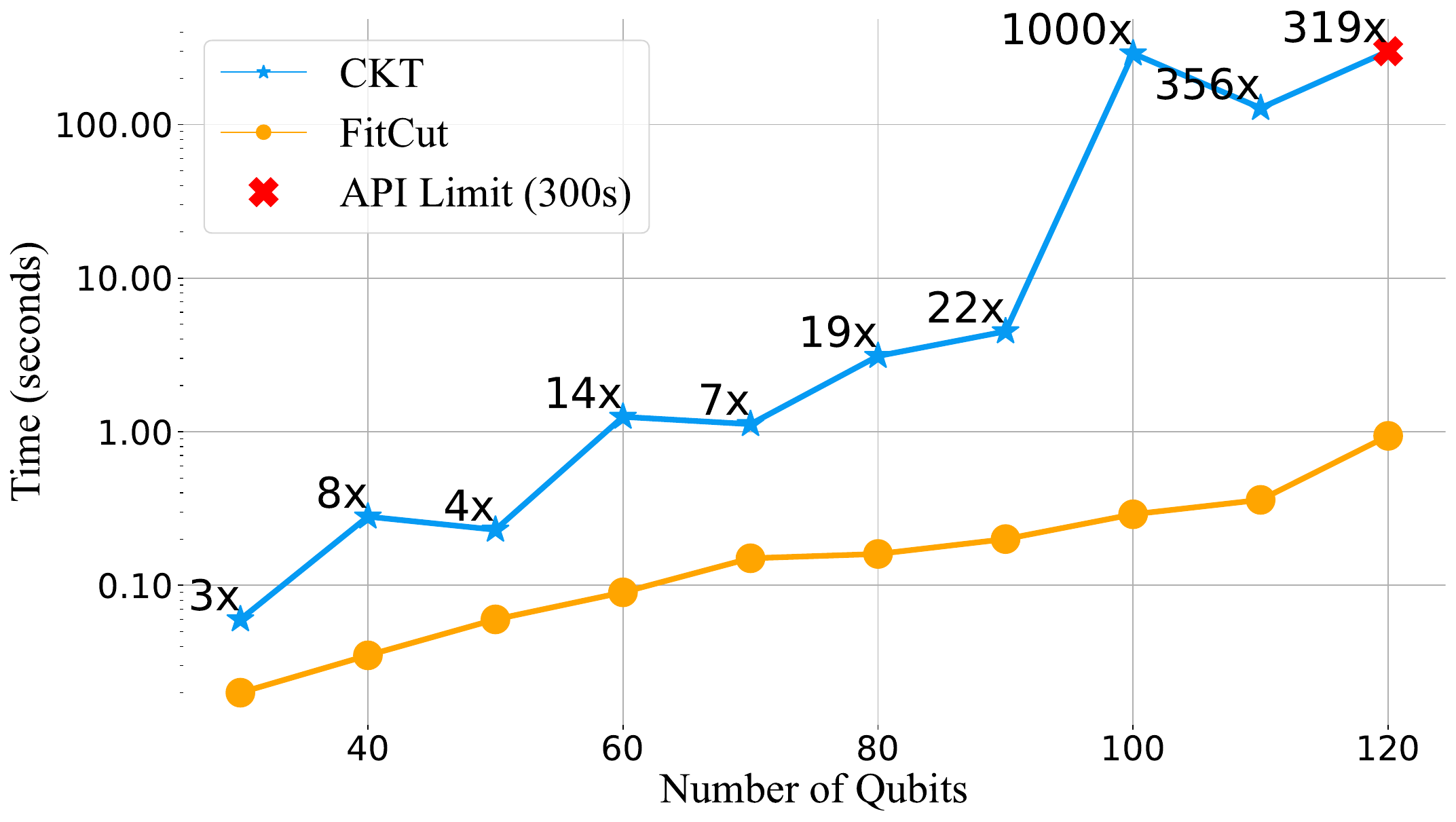}
        \caption{BV}
        \label{fig:bv_run_20}
    \end{subfigure}
  \hfill
    \begin{subfigure}[b]{0.23\textwidth}
        \includegraphics[width=\textwidth]{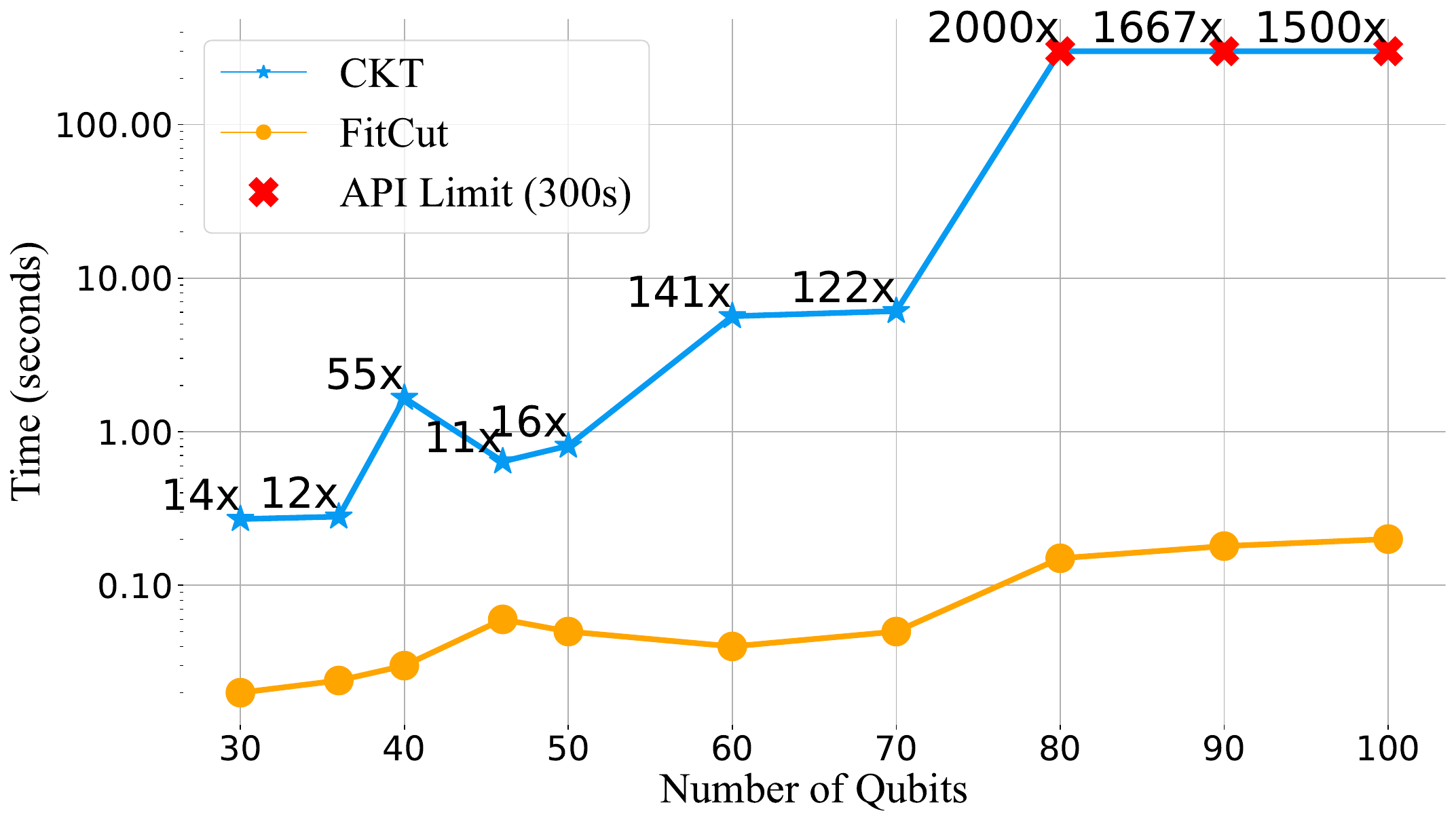}
        \caption{HWEA}
        \label{fig:hwea_run_20}
    \end{subfigure}
  \hfill  
    \begin{subfigure}[b]{0.23\textwidth}
        \includegraphics[width=\textwidth]{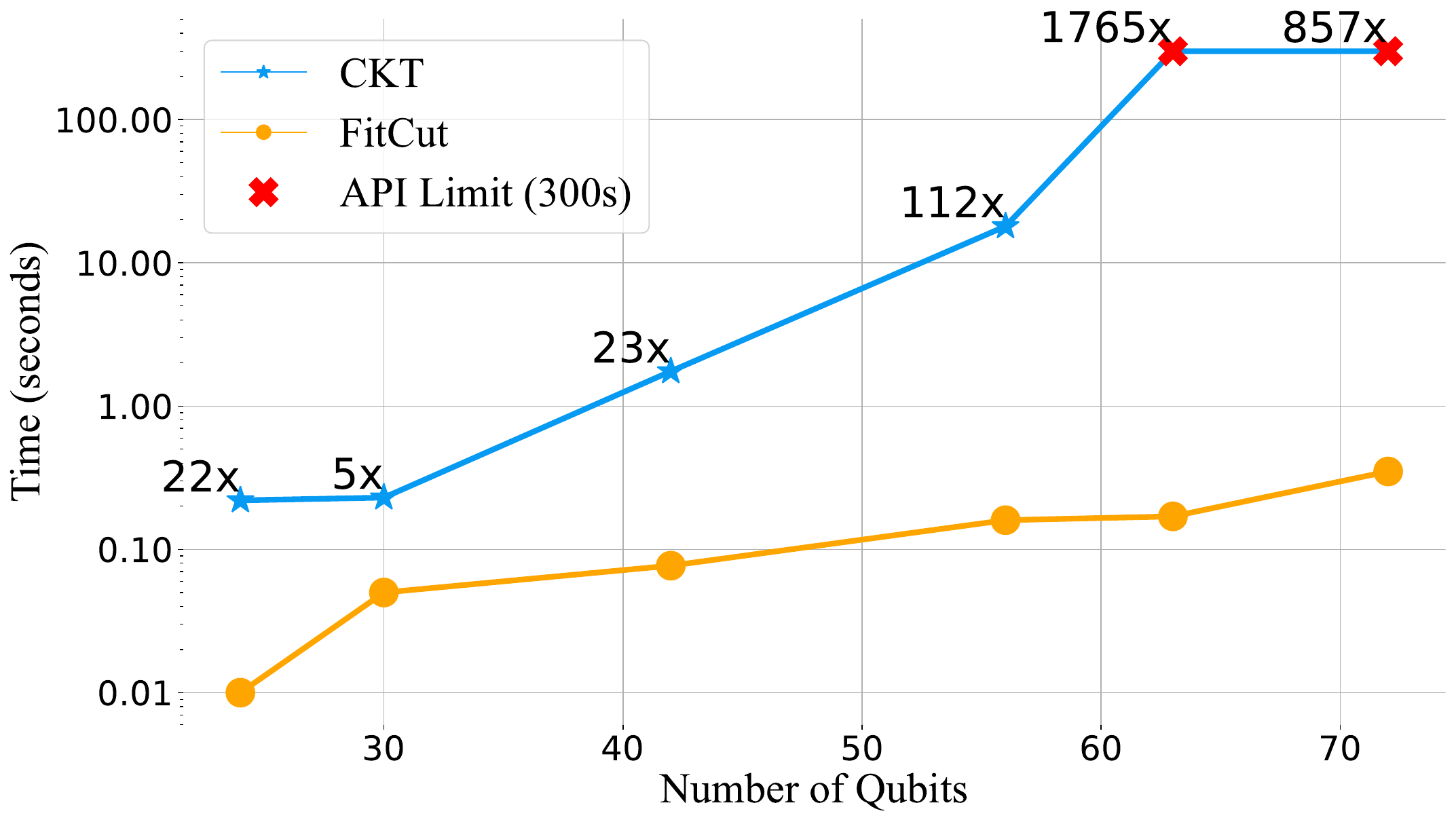}
        \caption{Supremacy}
        \label{fig:supremacy_run_20}
    \end{subfigure}
\caption{Runtime comparison in worker with 20-qubit}
\label{fig:worker20}
\end{figure*}

\subsection{Runtime}

We conducted two sets of experiments to evaluate the runtime performance of our system, specifically focusing on the cutting time cost. The performance is influenced by two main factors: (1) the qubit count of input circuits and (2) the qubit constraints on quantum workers. For instance, the Adder circuit was tested in 13 variants with qubit counts ranging from 20 to 80, increasing in increments of 6 qubits, while the BV circuit had 10 variants with counts increasing from 30 to 120 in increments of 10 qubits. Besides, \sol~ runs 50 times for each experiment and record the average time cost. 

Additionally, we assessed the impact of resource constraints on quantum workers, evaluating configurations with 15 and 20 qubits. Naturally, a lower qubit constraint leads to a higher number of required subcircuits, impacting overall system performance. Another underlying factor is the number of two-qubit gates in the input circuits. An increase in the number of two-qubit gates results in a corresponding increase in the number of nodes and edges in both $DAG$ and \sol~ weighted graph. For instance, a 30-qubit Adder circuit contains 239 nodes and 448 edges, whereas a 30-qubit BV circuit comprises 29 nodes and 28 edges. Consequently, the computational effort required for the CKT searcher to find partitions for a 15-qubit worker is significantly higher for the Adder circuit, taking approximately 4 seconds, compared to approximately 0.2 seconds for the BV circuit.


In Figure \ref{fig:worker15} and Figure \ref{fig:worker20}, we present the runtime results of \sol~ and CKT. Besides lines, the red cross means CKT API reaches its upper limit, 300 seconds, and fails to return any solutions. Additionally, the numbers on top of the blue lines present the order of improvements. 
It is evident that \sol~ significantly outperforms CKT across all test scenarios by several orders of magnitude. For instance, under a 15-qubit constraint, the runtimes for various Adder circuit variants with \sol~ are 0.05, 0.06, 0.12, 0.12, 0.15, 0.21, 0.32, 0.41, and 0.52 seconds, compared to CKT's 3.61, 4.23, 12.58, 26.09, 300, 99.54, 86.92, 300, and 300 seconds. Remarkably, \sol~ improves runtime efficiency by 67x to 2000x. Notably, CKT's API has a runtime limit of 300 seconds; values reaching this limit indicate the search has been terminated early and if available, the best current solution will be returned. It's important to note that no global optimal solution is guaranteed in such cases. Specifically, \sol~ outperforms CKT by more than 2000x in the scenario involving a 50-qubit Adder circuit under a 15-qubit worker constraint. This substantial performance gain is achieved because \sol~ optimizes the search process by merging communities in consideration of resource constraints, aiming to find a suboptimal solution that maximizes resource utilization, rather than enumerating all possible cutting points in the search space.

\begin{table}[ht]
	\centering
	\caption{Cut Numbers}
	\scalebox{0.75}{
            \begin{tabular}{|c|c|c|c|c|c|c|}
        \hline
        \multirow{2}{*}{circuit type}  & \multicolumn{3}{c|}{15-qubit Constraint} & \multicolumn{3}{c|}{20-qubit Constraint} \\
        \cline{2-7}
        &  Qubit Count & CKT & FitCut & Circuit Qubit & CKT & FitCut \\
        \hline
        
        \multirow{6}{*}{Adder} & 20 & 2 & 2 & 30 & 2 & 2 \\
        \cline{2-7}
        & 30 & 4 & 4 & 40 & 4 & 4 \\
        \cline{2-7}
        & 40 & 6* & 6 & 50 & 4 & 4 \\
        \cline{2-7}
        & 50 & 6 & 6 & 60 & 6 & 6 \\
        \cline{2-7}
        & 54 & 8* & 8 & 70 & 6 & 6 \\
        \cline{2-7}
        & 60 & NA & 8 & 80 & NA & 8 \\
        \hline
        
        \multirow{6}{*}{BV} & 50 & 3 & 3 & 30 & 1 & 1 \\
        \cline{2-7}
        & 60 & 4 & 4 & 50 & 2 & 2 \\
        \cline{2-7}
        & 70 & 4 & 4 & 70 & 3 & 3 \\
        \cline{2-7}
        & 80 & 5 & 5 & 90 & 4 & 4 \\
        \cline{2-7}
        & 90 & 6* & 6 & 110 & 5 & 5 \\
        \cline{2-7}
        & 100 & 7* & [7,8] & 120 & 6* & 6 \\
        \hline

        \multirow{6}{*}{HWEA} & 20 & 2 & 2 & 30 & 2 & 2 \\
        \cline{2-7}
        & 30 & 4 & 4 & 40 & 4 & 4 \\
        \cline{2-7}
        & 40 & 6 & [4,6] & 50 & 4 & 4 \\
        \cline{2-7}
        & 50 & 6 & [6,8] & 60 & 6 & 6 \\
        \cline{2-7}
        & 60 & 8 & [8,11] & 70 & 6 & [6,8] \\
        \cline{2-7}
        & 70 & 10* & [10,14] & 80 & 8* & [8,10] \\
        \hline

        \multirow{6}{*}{Supremacy} & 20 & 4 & [5,8] & 24 & 4 & 2 \\
        \cline{2-7}
        & 30 & 8 & [12,15] & 30 & 5 & [6,9] \\
        \cline{2-7}
        & 36 & 11 & [12,18] & 42 & 10 & [10,15] \\
        \cline{2-7}
        & 42 & 13 & [14,15] & 56 & 15 & [16,22] \\
        \cline{2-7}
        & 49 & 21* & [17,23] & 63 & 20* & [20,26] \\
        \cline{2-7}
        & 56 & 25* & [22,28] & 72 & 27* & [24,30] \\
        \hline
        \multirow{2}{*}{Note:} & \multicolumn{6}{|c|}{\textit{\#* : CKT terminates early due to time limit (300s)}} \\
        & \multicolumn{6}{|c|}{\textit{NA : CKT can't find result}} \\
        \hline
        \end{tabular}
        }
\label{table:Cut-number}
\vspace{-0.15in}
\end{table}


A consistent trend emerges across all tested experiments. Specifically, CKT manages to return optimal solutions within a reasonable timeframe when the input circuits are smaller, attributable to their limited search spaces. For example, with a 20-qubit constraint, the runtime for the smallest-sized Adder, BV, HWEA, and Supremacy circuits is 0.22, 0.06, 0.33, and 0.81 seconds, respectively. In comparison, \sol~ significantly outperforms CKT, clocking in at 0.01, 0.02, 0.02, and 0.06 seconds respectively, yielding improvements of 22x, 3x, 16.5x, and 13.5x. As the size of the input circuits increases, \sol~ demonstrates far greater stability than CKT. For instance, as the BV circuit size expands from 30 to 120 qubits under a 20-qubit constraint, CKT's runtime escalates from 0.06 to 300 seconds (reaching the upper limit) and fails to return a solution. Conversely, \sol's time cost only increases from 0.02 to 0.94 seconds, merely 47 times, and stays within a reasonable timeframe. This indicates that while \sol~ focuses on finding suboptimal solutions, it is substantially more scalable than CKT.

\begin{figure*}[htbp]
    \centering
\begin{subfigure}[b]{0.45\textwidth}
  \centering
  \includegraphics[width=1\linewidth]{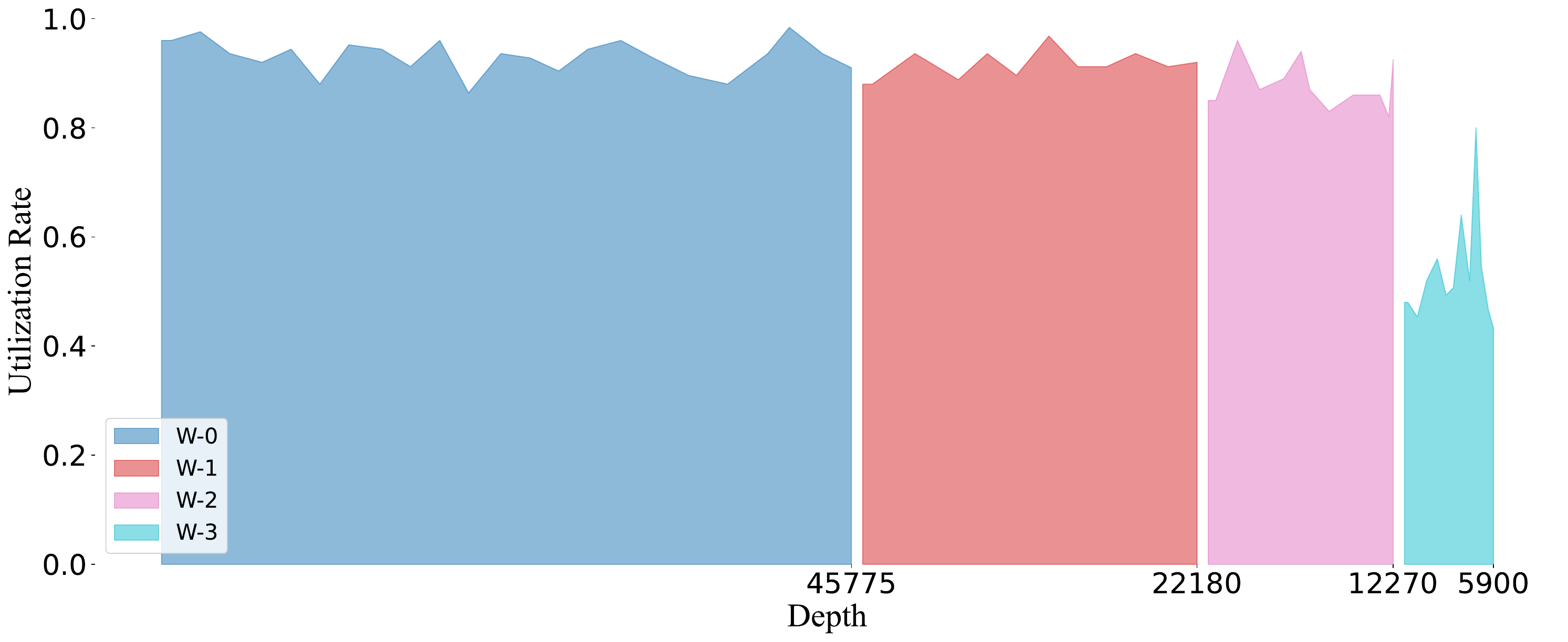}
    \caption{FitCut}
    \label{fig:FitCut}
\end{subfigure}
~
\begin{subfigure}[b]{0.45\textwidth}
  \centering
  \includegraphics[width=1\linewidth]{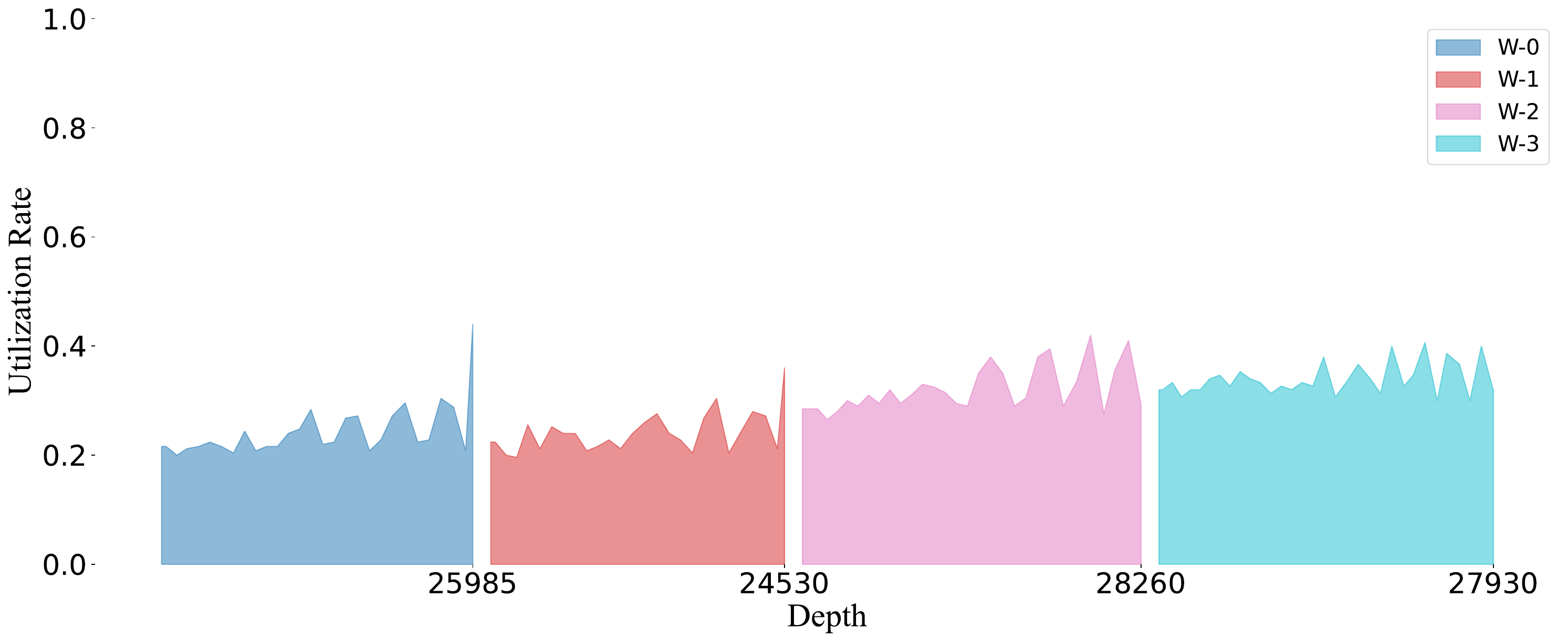}
    \caption{Modularity-only Solution}
    \label{fig:baseline}
\end{subfigure}
\caption{Depth-based Resource Utilization in a 4-worker setting: 25, 25, 20, 15 qubits on W-1, W-2, W-3, W-4. }
\label{fig:cluster}
\end{figure*}


Our study of the effects of different resource constraints reveals that \sol~ exhibits greater elasticity compared to CKT. For instance, with the HWEA circuit when the input size is 60 qubits, CKT's runtime is 31.02 seconds under a 15-qubit constraint and 5.65 seconds under a 20-qubit constraint, yielding a 5.48x difference in performance. In contrast, \sol~ records runtimes of 0.07 seconds and 0.04 seconds under the same constraints, showing a less pronounced 1.75x difference. Additionally, for a 42-qubit Supremacy circuit, the disparity in performance scales down to 5.14x for CKT and 2.59x for \sol, further demonstrating \sol's superior adaptability to varying resource constraints.

\subsection{Number of Cuts}

CKT aims to minimize the number of cuts under a given qubit constraint, while \sol~ employs a two-tiered objective function designed to maximize utilization rates without compromising cutting costs. As part of \sol's community-based optimization strategy, \sol~ avoids further dividing nodes that are initially clustered into the same community. Although this approach may result in a higher number of cuts, providing a suboptimal solution, it significantly reduces the search space. Table~\ref{table:Cut-number} displays the number of cuts under various scenarios. We executed \sol~ 50 times, recording the range of cuts when multiple outcomes occurred. For scenarios where CKT failed to return a solution within the 300-second limit, values are marked with a star (*).



From the data in Table~\ref{table:Cut-number}, it is evident that Adder variants perform identically to CKT, both returning optimal solutions in all completed scenarios. In complex cases such as 60-qubit input circuits with a 15-qubit constraint and 80-qubit input circuits with a 20-qubit constraint, CKT failed to find feasible solutions within the allocated 300-second limit. \sol~ consistently achieves a minimal number of cuts in all 50 experiments for Adder and BV. For Supremacy, CKT hits the 300-second limit and returns the best current solution for the 49-qubit and 56-qubit input circuits with a 15-qubit constraint. In these cases, \sol~ achieves fewer number of cuts with 800X to 1000X speedup. For HWEA, \sol~ completes the search quickly but shows variability in cut numbers. For instance, in the 70-qubit HWEA scenario, \sol~ reports cut numbers ranging from 10 to 14, with 10 being the optimal number. This variability arises from the non-deterministic nature of the community detection step, where \sol~ randomly shuffles nodes in each iteration to maximize modularity (a local optimal). In subsequent stages, \sol~ focuses on merging communities to optimize resource utilization under predefined constraints. Overall, \sol~ maintains reasonable and stable performance across 50 experiments under various settings.

\subsection{Resource Utilization}
\sol~ employs a two-tiered objective function that optimizes both cutting efficiency and resource utilization. We use a depth-based utilization metric that accounts for both the width (number of qubits) and depth (maximum number of gates) of the circuits executed. For a circuit with width $i$ and depth $j$ being executed on a worker with $m$ qubits, the qubit utilization is calculated as $\frac{i}{m}$, lasting for $j$ depth units.

Figure~\ref{fig:cluster} illustrates resource utilization in a setting with 4 quantum workers with different qubit capacities: 25 qubits for W-1 and W-2, 20 qubits for W-3, and 15 qubits for W-4. The Y-axis represents the qubit utilization rate, and the X-axis shows the depth. \sol~significantly enhances qubit utilization across all workers. For instance, on W-1, \sol~achieves an average resource utilization rate of 0.93, compared to only 0.24 with a Modularity-only solution, marking a 3.88x improvement. Similar trends are observed with other workers, achieving 3.79x, 2.75x, and 1.56x improvements for W-2, W-3, and W-4, respectively. This increase is attributed to \sol~considering each worker's qubit capacity during the circuit cutting and merging process, aiming to optimize the fit of subcircuits rather than merely minimizing their width. Overall, \sol~achieves a system-wide utilization rate of 0.83, a 2.86x improvement over the 0.29 rate of the Modularity-only approach. Notice that the accumulated depth of executed circuits is not evenly distributed among the workers. \sol~prioritizes resource utilization over depth distribution, and W-1 and W-2 are assigned more circuits due to their larger capacities, which reduces the number of necessary cuts and enhances system efficiency and scalability. However, when comparing the system-wide accumulated depth, \sol~reduces 19.3\% as it utilizes more qubits, 86,125 vs 106,705.

%% file: conclusion.tex
\section{Discussion and Conclusion}

In this paper, we have introduced \sol, a scalable circuit cutting and scheduling system designed for multi-node, distributed quantum systems. \sol~transforms a circuit’s $DAG$ representation into a gate-only weighted graph. Employing a community-based, bottom-up approach, \sol~efficiently cuts large circuits into smaller subcircuits tailored to the qubit capacities of quantum workers. It also schedules these subcircuits by maximizing the resource utilization rates of each worker. Implemented with Qiskit and evaluated extensively, \sol~demonstrates a significant reduction in cutting time costs, achieving improvements ranging from 3 to 2000 times compared to the Qiskit CKT. Moreover, \sol~enhances resource utilization rates by up to 3.88 times per worker and 2.86 times across the system. These results underscore \sol's effectiveness in optimizing the performance of distributed quantum computing environments.

\section{Acknowledgement}
 This research was supported in part by the National Science Foundation (NSF) under grant agreements 2329020, 2301884 and 2335788.
This work was also partially supported by the U.S. Department of Energy, Office of Science, National Quantum Information Science Research Centers, Quantum Science Center. This research used resources of the Oak Ridge Leadership Computing Facility, which is a DOE Office of Science User Facility supported under Contract DE-AC05-00OR22725. This research used resources of the National Energy Research Scientific Computing Center (NERSC), a U.S. Department of Energy Office of Science User Facility located at Lawrence Berkeley National Laboratory, operated under Contract No. DE-AC02-05CH11231.